\documentclass{article}
\usepackage{graphicx}  
\usepackage{amsmath}   
\usepackage[compress]{cite}
\usepackage{amssymb}   
\usepackage{bm} 
\usepackage{dcolumn}
\usepackage{color}
\usepackage{mathrsfs}
\usepackage{amsfonts}
\usepackage{varioref}
\usepackage{braket}
\usepackage{subcaption}
\usepackage{comment}
\RequirePackage[colorlinks,citecolor=blue,urlcolor=magenta,linkcolor=blue]{hyperref}
\labelformat{section}{Section #1} 
\labelformat{subsection}{Section #1} 
\labelformat{subsubsection}{Section #1}
\labelformat{subsubsubsection}{Section #1}
\labelformat{equation}{Eq.~(#1)} 
\labelformat{figure}{Fig.~#1} 
\labelformat{subfigure}{Fig.~\thefigure#1} 
\labelformat{table}{Tab.~#1} 
\labelformat{appendix}{Appendix #1}
\title{Effective-one-body formalism for leading-order radiative effects in the post-linear framework }
\author{Karthik Rajeev\footnote{karthik\_rajeev@iitb.ac.in}\quad and\quad  S. Shankaranarayanan\footnote{shanki@iitb.ac.in}
	\\
	{\small{Department of Physics}}\\
		{\small{Indian Institute of Technology Bombay, Mumbai, 400076, India}}}
\date{ }  

\begin{document}
	\maketitle
\begin{abstract}
In recent years, significant progress has been made in the computation of conservative and dissipative scattering observables using the post-Minkowskian approach to gravitational dynamics. However, for accurate modelling of unbound orbits, an appropriate effective-one-body (EOB) resummation of the post-Minkowski results that also accounts for dissipative dynamics is desirable. As a step in this direction, we consider the electromagnetic analog of this problem here. We show that a six-parameter equation of motion encapsulates the effective-one-body dynamics for the electromagnetic scattering problem appropriate to third-order in the coupling constant. Three of these six parameters describe the conservative part of the dynamics, while the rest correspond to the radiation-reaction effects. Here we show that only two radiation-reaction-related parameters are important at the desired order, making the effective number of parameters in our formalism to be five. We compute the explicit forms of these five parameters by matching EOB scattering observables to that of the original two-body ones computed by Saketh et al. ~\cite{Saketh:2021sri}. Interestingly, our formalism leads to a conjecture for the sub-leading angular momentum loss, for which no precise computations exist. In addition, we demonstrate that the bound-orbit observables computed using our method are in perfect agreement with those calculated using unbound-to-bound analytical continuation techniques. Lastly, we qualitatively discuss the extension of our formalism to gravity.
\end{abstract}

\section{Introduction}
With the dawn of gravitational wave (GW) astronomy\cite{LIGOScientific:2016aoc,LIGOScientific:2016dsl,LIGOScientific:2017vwq,LIGOScientific:2018mvr,LIGOScientific:2020ibl}, a new avenue has opened up to explore the deeper aspects of astrophysics and fundamental physics. Besides, the gravitational wave (GW) detectors' network is expanding and being upgraded at a promising pace\cite{Punturo:2010zz,2017arXiv170200786A,Reitze:2019iox}. Future gravitational wave detectors, like Cosmic explorer and Einstein telescope, will have a very high signal-to-noise ratio and potentially help test General Relativity (GR) in strong-gravity regimes~\cite{Evans:2016mbw,Barack:2018yly}. This, in turn, has led to an increasing demand for accurate theoretical modeling of binary systems in  GR and beyond. Among the several complementary theoretical perspectives on the two-body problem in relativistic gravity, the study of scattering dynamics has recently attracted much attention.   

The post-Minkowskian (PM) scheme~\cite{PhysRevD.102.024060,PhysRevD.94.104015,PhysRevD.97.044038,Buonanno:1998gg}, wherein one computes the classical scattering observables in a weak field expansion and without restriction on velocities, has benefited from the recent progress made in scattering amplitude computations and other quantum field theory techniques \cite{Amplitude1,Amplitude2,Neill:2013wsa,Amplitude3,Amplitude4,Amplitude5,Amplitude6,Amplitude7,Amplitude8,Amplitude9,Amplitude10,Amplitude11,Amplitude12,Amplitude13,Amplitude14,Amplitude15,Amplitude16,Amplitude17,Amplitude18,Amplitude19,Amplitude20,Amplitude21,Amplitude22,Amplitude23,Amplitude24,Amplitude25,Amplitude26}. 
Although PM formalism is naturally adapted for two-body gravitational scattering, one can also extract information about bound binary systems from the scattering results by mapping to effective-one-body (EOB) \cite{Buonanno:1998gg,PhysRevD.94.104015,PhysRevD.97.044038}, effective field theory (EFT) methods~\cite{Neill:2013wsa,Amplitude2} and judicious analytic continuation prescriptions~\cite{Amplitude7,Kalin:2019inp}. Therefore, in particular, the PM scheme, in conjunction with EOB, furnishes a promising theoretical framework to address the full binary dynamics in GR.    

Initially formulated for the post-Newtonian (PN) framework, the EOB formalism in GR has been extended only recently for the PM approach\cite{PhysRevD.94.104015}. A variant of the EOB formalism for PM that is more compatible with scattering-amplitude-based techniques has also been developed~\cite{Damgaard:2021rnk}. The vast majority of PM literature until now, however, has focused on the conservative sector of the binary gravitational dynamics. The same is true about the literature on EOB re-summation of PM results. However, dissipation and radiation are key features of binary systems. Therefore, it is imperative to have a better theoretical understanding of the dissipative effects in such systems. However, there have been promising developments in this field in recent years~\cite{Manohar:2022dea,Amplitude10,Dlapa:2022lmu,Damour:2020tta,Amplitude13,Cho:2021arx,Amplitude18,Amplitude22,Bini:2021qvf} that are pretty promising. Important progress can be made by constructing an EOB formalism that systematically accounts for dissipative and conservative effects at the necessary order in PM approximation and naturally connects to scattering amplitude-based methods. 

In light of this, it is instructive to look at a closely related but more easily tractable problem, namely, to set up the EOB formulation of binary scattering dynamics in electromagnetism (EM). The scattering of charged particles in EM and, more generally, Yang-Mills theories is of significant interest. Although the difficulties introduced by non-linearities are absent, the classical scattering problem in the Abelian gauge theory shares some of the technical hurdles one runs into in GR \cite{Saketh:2021sri}. Moreover, the more direct gauge theory results may also be used to build the corresponding gravity results by using an appropriate version of the double-copy relations~\cite{Kawai:1985xq,Bern:2008qj,Bern:2010ue,Bern:2019prr}. Hence, the EOB formalism for EM theory, which is the focus of this paper, not only serves as a useful toy model but also potentially as a source of ingredients to construct the gravity results. While drafting this paper, an EOB formalism for conservative and dissipative PM dynamics was put forward in \cite{Damour:2022ybd}. The approach we set out to describe here is different in spirit from the one described in \cite{Damour:2022ybd} and, therefore, we hope that the same can, in turn, motivate a novel approach to EOB for dissipative PM dynamics. 

The paper is organized as follows: We review the classical electromagnetic scattering in \ref{review_scattering}. Then in \ref{EOB_setup}, we describe the effective-one-body formalism adapted to address the relativistic EM scattering dynamics. The formalism is concluded in \ref{EOB_conclude}, where we explicitly solve for the unknown coefficients in the EOB ansatz. In \ref{review_comparison}, we also present a brief comparison of our approach with certain other ones. Then, in \ref{unbound_to_bound}, we illustrate the application of our formalism for extracting results concerning bound orbits. Finally, we conclude with a summary and outline the future outlook in \ref{summary}.  

\textit{Notations and conventions:} Throughout the paper, we shall follow $(+,-,-,-)$ signature, $c=1$ and the notation $k\equiv 1/(4\pi\epsilon_0)$, where $\epsilon_0$ is the vacuum permittivity. We also follow the notations: $A.B\equiv A^{\mu}B_{\mu}$, where $A^{\mu}$ and $B^{\mu}$ are two four-vectors and $|V|^2\equiv-V.V$, for a space-like vector $V^{\mu}$.

\section{Review of 3rd order post-linear(PL) results}\label{review_scattering}

The post-Linear (PL) approach is analogous to the PM scheme, where one seeks a formal series solution to the classical scattering problem in gauge theory. Like the PM scheme, the series is expanded in coupling constant without any restriction on the velocities of the scatterers. For example, in Ref.~\cite{Westpfahl:1985tsl}, the scattering up to the second-order in the coupling (2PL) was performed for EM (and GR). Recently, Saketh et al. ~\cite {Saketh:2021sri} extended Westpfahl's results for EM to the third-order scattering problem, which we shall refer to as 3PL. As mentioned in the introduction, the key result of this work is a prescription for a convenient EOB re-summation of the EM scattering results up to 3PL and including leading order dissipative effects due to radiation reaction (RR). As we show, our approach is also efficient in providing a one-to-one correspondence between certain observables concerning bound and unbound orbits. Before discussing the EOB formalism, we briefly review 3PL results, closely following reference \cite{Saketh:2021sri}. 

\subsection{Kinematics}

We consider relativistic EM scattering of two particles of rest masses $m_{i}$ and charges $q_{i}$, respectively, where $i=1,2$. We solve the scattering problem as a function of the asymptotic initial momenta $p_{i}^{\mu}$ and the asymptotic initial impact parameter vector $b^{\mu}$. The initial momenta, in turn, can be written in terms of the initial velocities $u^{\mu}_{i}$ as $p^{\mu}_{i}=m_{i}u^{\mu}_{i}$. By definition, $b.u_{i}=0$, for $i=1,2$. 
While most of the results do not depend on the sign of $q_1 \times q_2$, for concreteness, we will assume $q_1q_2 < 0$. This will allow us to connect some EM results to gravitational results.

It is often convenient to rephrase these initial data in terms of certain center-of-momentum (COM) and `relative' quantities. The standard definitions for the initial relative velocity $v$, the corresponding Lorentz factor $\gamma$ and the COM energy $E$ are:
\begin{align}
	\frac{1}{\sqrt{1-v^2}}&=\gamma =\frac{p_1.p_2}{m_1 m_2}, \\\label{totE}
	E^2&=(p_1+p_2)^2=m_1^2+m_2^2+2m_1m_2\gamma=M^2+2mM(\gamma-1),
\end{align}
where, the total mass $M=m_1+m_2$ and the reduced mass $m = m_1m_2/M$. In addition, the COM velocity $U^{\mu}$ and the `relative momentum' $P^{\mu}$ are defined as:
\begin{align}
	U^{\mu}&=\frac{(p_1^{\mu}+p_2^{\mu})}{E},\\\label{relmom}
	P^{\mu}&=\frac{1}{E}\left[(p_2.U) p^{\mu}_1-(p_1.U) p_{2}^{\mu}\right],\\
	|P|&\equiv\left(\frac{M}{E}\right)(m\gamma v),
\end{align}
where, $|P|$ is the magnitude of the space-like vector $P^{\mu}$. With the above definitions, we find that the magnitude of initial COM angular momentum, $J$, has the form:
\begin{align}
	J= |P||b|=\left(\frac{M}{E}\right) \left(m\gamma v b\right).\label{angmom}
\end{align} 

\subsection{PL scattering dynamics}

The scattering dynamics is described in terms of the worldlines $z_{i}^{\mu}(\tau_i)$, parametrized by the proper-time $\tau_i$ of the respective particle. Further, we can describe the electromagnetic field generated by the particles by the gauge potential $A^{\mu}$. Assuming the Lorentz-gauge condition, $\partial_{\mu}A^{\mu}=0$, the worldlines are solutions to the following set of coupled differential equations:
\begin{align}\label{MaxwellEq}
	\Box A^{\mu}(x)&=4\pi k J^{\mu}(x)\equiv 4\pi\left[ \sum_{i=1}^{2}q_i\int d\tau_i\dot{z}^{\mu}_{i}\delta^4(x-z_{i})\right]\qquad&&\textrm{(Maxwell's Eqn.)}, \\\label{ADL}
	m_i \ddot{z}_i&=q_i F^{\mu}_{~~ \nu}(z_i)\dot{z}_i^{\nu}+\frac{2k q_i^2}{3}\left(\dddot{z}^{\mu}_i+\dot{z}^{\mu}_i \ddot{z}_{i}.\ddot{z}_{i}\right)\qquad &&\textrm{(Lorentz-Dirac Eqn.)},
\end{align}
where $F_{\mu\nu}=(\partial_{\mu}A_{\nu}-\partial_{\nu}A_{\mu})$ denotes the external field acting on a given particle, and the effects of the self-field are encapsulated in the second term in the RHS of \ref{ADL}. The exact scattering worldline functions $z^{\mu}_{i}(\tau_i)$ self-consistently satisfy \ref{MaxwellEq} and \ref{ADL}, with the initial conditions $\dot{z}^{\mu}_{i}(-\infty)=u^{\mu}_i$, $\lim_{\tau_{1}\rightarrow-\infty}z_1(\tau_1).b=b^2$, $\lim_{\tau_{2}\rightarrow-\infty}z_2(\tau_2).b=0$. 

However, even in the EM case, the exact solutions $z_{i}^{\mu}(\tau_i)$ are not available. Therefore, in the PL approach, one resorts to an iteration scheme, wherein one starts by expanding the worldlines as a series in coupling constant $k$\footnote{Technically, the series expansion is about a dimensionless quantity $\tilde{k}\equiv kq_1q_2/J$. However, we can be sloppy about it if we are sufficiently careful about the regime of validity ($\tilde{k}\ll 1$) of the results.}:
\begin{align}
	z_{1}^{\mu}(\tau_1)&=b^{\mu}+u^{\mu}_1\tau_1+k z^{(1)\mu}_{1}(\tau_1)+k^2 z^{(2)\mu}_{1}(\tau_1)+k^3z^{(3)\mu}_{1}(\tau_1)+...,\\
	z_{2}^{\mu}(\tau_2)&=u^{\mu}_2\tau_2+k z^{(1)\mu}_{2}(\tau_2)+k^2 z^{(2)\mu}_{2}(\tau_2)+k^3z^{(3)\mu}_{2}(\tau_2)+...
\end{align} 
Note that the zeroth-order solutions describe the free straight-line trajectories. The above series ansatz is then fed into the \ref{MaxwellEq} and \ref{ADL}, to iteratively solve the corrections to the straight line paths at progressively higher orders in $k$; hence the name post-linear formalism. In Ref. ~\cite{Saketh:2021sri}, the solution of the worldlines up to $z^{(2)}_{i}(\tau_i)$ and momenta up to $m_i \dot{z}^{(3)}_{i}(\tau_i)$ were explicitly computed and, from them, the relevant scattering observables were also extracted. Unfortunately, the precise forms of the 3PL worldlines are rather lengthy and not quite illuminating. Hence, in the next subsection, we shall briefly summarize the scattering observables presented in Ref.~\cite{Saketh:2021sri}.

\subsection{Scattering observables}

In this subsection, we concentrate on three scattering observables --- 
scattering angle, $\chi$, magnitude of the radiated angular momentum, $\delta J$ and  the radiated COM energy, $\delta E$ ---
computed in Ref.~\cite{Saketh:2021sri}, which are relevant to the current work. 
The scattering angle $\chi$ is defined via:
\begin{align}
	\sin\chi=\frac{-\Delta p_1. b}{|b||P(\infty)|}=\frac{-\Delta p_1. b}{|b||P|}+\mathcal{O}(k^4),
\end{align}
where, $\Delta p^{\mu}_1=m_1 \dot{z}_1(\infty)-m_1\dot{z}_1(-\infty)$ is the net impulse on particle-1 and $|P(\infty)|$ is the final relative momentum. The 3PL result for $\chi$ is~\cite{Saketh:2021sri}:
\begin{align}\label{chi_tot}
	\chi&=\chi_{\rm cons}+\chi_{\rm rad},\\\label{chi_cons}
	\chi_{\rm cons}&=\frac{2k q_1 q_2}{m\gamma|b|v^2}h_{\nu}(\gamma)-\frac{\pi k^2q_1^2q_2^2}{2m^2\gamma^2|b|^2v^2}h_{\nu}(\gamma)\\ \nonumber
	&+\frac{2k^3q_1^3 q_2^3\left[(2\gamma^2-3)h^2_{\nu}(\gamma)-6\nu v^4\gamma^3\right]}{3 m^3|b|^3\gamma^5v^6}h_{\nu}(\gamma)	+\mathcal{O}(k^4),\\\label{chi_rad}
	\chi_{\rm rad}&=-\frac{4k^3q_1^2q_2^2}{3m|b|^3\gamma v^3}\left[\left(\frac{q_1^2}{m_1^2}+\frac{q_2^2}{m_2^2}\right)-\frac{3q_1 q_2}{m_1 m_2}\left(\frac{1}{\gamma v^2}-\frac{\tanh^{-1}(v)}{\gamma^3v^3}\right)\right]h_{\nu}(\gamma)+\mathcal{O}(k^4),
\end{align}
where $\chi_{\rm cons}$($\chi_{\rm rad}$) denotes the conservative (radiative) part of the scattering angle. The symmetric mass ratio ($\nu$) and $h_{\nu}(\gamma)$ are defined as:  
\begin{equation}
\label{def:nu-hnu}
\nu=\frac{m}{M}; \qquad 
h_{\nu}(\gamma)= \frac{E}{M} =\sqrt{1+2\nu(\gamma-1)} \, .   
\end{equation}
The radiated angular momentum $\delta J$ is the negative of the binary system's net change in total angular momentum. Hence, $\delta J=J_{i}-J_{f}$, where $J_{i}$ and $J_{f}$ are the initial and final angular momenta, respectively. With this definition, the radiated angular momentum in the COM frame $\delta J$ reduces to:
\begin{align}\label{delJ}
	\delta J&\equiv k^2\delta J_2+k^3\delta J_{3}+...\\
 &=-\frac{4 k^2 q_1 q_2 m\gamma}{3|b|h_{\nu}(\gamma)}\left[\left(\frac{q_1^2}{m_1^2}+\frac{q_2^2}{m_2^2}\right)-\frac{3q_1 q_2}{m_1 m_2}\left(\frac{1}{\gamma v^2}-\frac{\tanh^{-1}(v)}{\gamma^3v^3}\right)\right]+\mathcal{O}(k^3).
\end{align}  

The radiated linear momentum $K^{\mu}$ is defined as the negative of the sum of impulses on both particles. Hence, $K^{\mu}=-\Delta p_1^{\mu}-\Delta p_2^{\mu}$, where $\Delta p_{i}^{\mu}=m_i \dot{z}^{\mu}_i(\infty)-p^{\mu}_i$, for $i=1,2$. The radiated COM energy $\delta E$ is then defines as $K^{\mu}U_{\mu}$ and reduces to:
\begin{align}\label{delE}
	\delta E&\equiv k^2\delta E_2+k^3\delta E_3+... \\
 &=\frac{\pi k^3q_1^2q_2^2}{4 |b|^3h_{\nu}(\gamma)}\left[\frac{3\gamma^2+1}{3\gamma v}\left(\frac{q_1^2}{m_1^2}+\frac{q_2^2}{m_2^2}\right)+\frac{(\gamma-1)(3\gamma^2+1)}{3\gamma v}\left(\frac{q_1^2m}{m_1^3}+\frac{q_2^2m}{m_2^3}\right)-\frac{\mathcal{G}(\gamma)}{\gamma v}\frac{q_1q_2}{m_1m_2}\right]+\mathcal{O}(k^4),\\
	\mathcal{G}(\gamma)&=\frac{(3\gamma^2+1)}{(\gamma v)^2}\left(\gamma-\frac{\tanh^{-1}(v)}{\gamma v}\right)-\frac{4}{(\gamma v)^2}(\gamma-1)^2.
\end{align}
As mentioned earlier, the main aim of this work is to devise an EOB formulation that, in addition to effectively capturing the perturbative information contained in the above two-body scattering observables, also encapsulates certain non-perturbative aspects. This latter feature, in turn, can potentially enable us to go beyond the scope of the weak field regime of PL and also gain insights into the bound binary dynamics. 

\section{Effective-one-body formalism: the set up}\label{EOB_setup}

The effective-one-body (EOB) formalism is the canonical transformation that maps the dynamics of a relativistically interacting two-body system to that of a single particle in an external background. More specifically, the EOB maps the compact binary dynamics of general relativity to the test particle motion in an external background metric. The EOB formalism was first introduced in the seminal paper~\cite{Buonanno:1998gg} and further developed in several works~\cite{Buonanno:2000ef,Damour:2000we,Damour:2001tu,Buonanno:2005xu}, is now routinely used in the detection of gravitational waves in LIGO-VIRGO-KARGA. 

Incidentally, the original paper of Buonanno and Damour cites an old work of Br\'{e}zin, Itzykson and  Zin-Justin (BIZ)\cite{PhysRevD.1.2349}, that essentially discusses an EOB formalism for EM, as one of the primary inspirations. The BIZ paper used an approximate summation of the ``crossed-ladder'' Feynman diagrams of $2\rightarrow2$ EM scattering amplitude, which essentially recovered the eikonal asymptotic behavior and arrived at the bound-state energy spectrum $E_n$ by inspecting the poles of the corresponding Green's function. Although this approximation did not effectively capture the centrifugal effects, their expression for $E_n$ correctly accounted for the recoil effects. The BIZ expression for the two-body bound state energy $E_n$ can be mapped to the relativistic \textit{one-body} spectrum $\epsilon_n$ of a particle, with mass being the reduced mass, moving in a static Coulomb potential as $E_n^2=m_1^2+m_2^2+2(m_1+m_2)\epsilon_n$. However, the BIZ paper does not account for \emph{the radiative effects}.

In the present discussion, we shall formulate an EOB approach to EM that is naturally adapted for the \textit{classical} PL scheme. In the rest of this section, we briefly discuss EOB basics for completeness. In the following sections, we explicitly show that our proposal efficiently accounts for the leading order radiation-reaction effects. We also qualitatively discuss how our proposal can be extended to GR.   

\subsection{EOB kinematics}\label{EOBkin}

As mentioned above, the starting point of EOB formalism is establishing a map between relevant quantities of the original 2-body problem and the effective one-body system. Due to the choice of canonical transformations, this mapping has some freedom. 
Here, we consider a choice introduced in \cite{Damgaard:2021rnk} in the context of EOB formalism for PM gravity. This choice allows the conservative EOB dynamics up to 3PM to be cast as the motion of a particle in an effective metric. In contrast, the conventional mapping used in Refs.~\cite{Buonanno:1998gg,Buonanno:2000ef} leads to modifying the standard mass-shell condition. Moreover, the new mapping introduced in \cite{Damgaard:2021rnk} also connects more directly with the results of the original 2-body scattering observables.  

The three key ingredients of the EOB mapping are: (i) energy map, (ii) momentum map, and (iii) angular momentum map. The total energy $E$, the relative momentum $P^{\mu}$ and the angular momentum $J$ for a system of free particles are defined in \ref{totE}, \ref{relmom} and \ref{angmom}, respectively. We shall denote the energy, spatial momentum, and angular momentum of the reduced mass in the effective-one-body description by $\epsilon$, $p^{\mu}$, and $j$, respectively. The EOB mapping amounts to the following identification:
\begin{align}
	\epsilon&=\frac{E^2-m_1^2-m_2^2}{2M},\\
	p^{\mu}&=h_{\nu}(\gamma)P^{\mu},\\
	j&=h_{\nu}(\gamma)J,\\
	|b|&\rightarrow |b|,\\
	\chi&\rightarrow \chi,
\end{align} 
where $h_{\nu}(\gamma)$ is defined in \ref{def:nu-hnu}. The last two equations above emphasize that the impact parameters and scattering angles of the original 2-body system and the EOB problem are identified. It is easily seen that $j=mv\gamma |b|$, $|p|=m v\gamma$ and $\epsilon=m\gamma$, as is desired. 

\subsection{EOB dynamics}\label{EOBdyn}

Let $x^{\mu}(\tau)=(t(\tau),\vec{x}(\tau))$ denote the effective worldline that describes the relative dynamics of the 2-body such that $m \dot{x}^{i}=p^{i}$, where the dot ($\dot{~}$) denotes derivative with respect to the proper-time of the reduced mass. The final ingredient of the EOB formalism for PL is a prescription for effective dynamics for $x^{\mu}$, which encapsulates all the information contained in the 2-body scattering observables up to the desired order. 

The success and utility of the conventional EOB formalism for GR stem from the fact that once the relevant parameters of the effective dynamics are fixed by matching the appropriate observables to a given order of an approximation scheme (like, for instance, PM, post-Newtonian), the EOB formalism can make sensible predictions even a bit beyond the regime of validity of the original approximation scheme. This feature can be attributed to the fact that the EOB formalism implicitly re-sums the approximate series expansion (i.e., either PM or PN expansions) and effectively translate the same to a systematic deformation about the test-particle limit (i.e., $\nu\rightarrow 0$). For instance, by construction, the effective metric in the PN EOB formalism can be viewed as a $\nu-$deformed Schwarzschild metric with the mass parameter being $M=m_1+m_2$ (see, for instance, \cite{Buonanno:1998gg,Buonanno:2000ef,Buonanno:2005xu}). 
 
Motivated by this, we seek an EOB dynamics for the PL formalism that can also be viewed as a deformation of the test-particle limit in EM. The equation of motion in the test-particle limit $m_1/m_2\rightarrow 0$ is given simply by the Lorentz-Dirac equation:
\begin{align}\label{ADL2}	m_1\ddot{x}^{\mu}=q_1F^{(c)\mu}_{\quad~~\nu}\dot{x}^{\nu}+\frac{2kq_1^2}{3}\left(\dddot{x}^{\mu}+\dot{x}^{\mu} \ddot{x}.\ddot{x}\right) \, ,
\end{align}
where $F^{(c)}_{\quad\mu\nu}$ is the static Coulomb field generated by the charge $q_2$. It is instructive to define the Coulomb force-field tensor $\mathcal{F}^{(c)}_{\quad\mu\nu}$ via $\mathcal{F}^{(c)}_{\quad\mu\nu}\equiv q_1 F^{(c)}_{\quad\mu\nu}$. The Coulomb force-field tensor can be written in terms of the vector potential $\mathcal{A}^{(c)}_{\quad\mu}$, which takes the form:
\begin{align}
	\mathcal{A}^{(c)}_{\quad\mu}=\left(\frac{k q_1q_2}{r},0,0,0\right).
\end{align}
The Lorentz-Dirac equation, recast in terms of the Coulomb force field tensor, is given by
\begin{align}\label{ADLmodified}	m_1\ddot{x}^{\mu}&=\mathcal{F}^{(c)\mu}_{\quad~~\nu}\dot{x}^{\nu}\\\nonumber
	&+\frac{2km_1}{3}\left[\left(\frac{q_1^2}{m_1^2}\right)\mathcal{F}^{(c)\mu}_{\quad~~\nu,\alpha}\dot{x}^{\nu}\dot{x}^{\alpha}+\left(\frac{q_1^2}{m_1^3}\right)\left(\mathcal{F}^{(c)\mu}_{\quad~~\nu}\mathcal{F}^{(c)\nu}_{\quad~~\alpha}\dot{x}^{\alpha}-\mathcal{F}^{(c)\alpha}_{\quad~~\nu}\mathcal{F}^{(c)\nu}_{\quad~~\beta}\dot{x}^{\beta}\dot{x}_{\alpha}\dot{x}^{\mu}\right)\right.\\\nonumber
	&\left.+\frac{2k}{3}\left(\frac{ q_1^4}{m_1^3}\right)\mathcal{F}^{(c)\mu}_{\quad~~\nu,\alpha\sigma}\dot{x}^{\nu}\dot{x}^{\alpha}\dot{x}^{\sigma}\right]+\mathcal{O}(k^4) \, .
\end{align}
We neglect the ${O}(k^4)$ terms since they will not be relevant for the 3PL results. Now, motivated by the form of the test-particle limit given by \ref{ADLmodified}, we propose that the EOB dynamics is described by the following deformed Lorentz-Dirac equation:
\begin{align}\label{ADLdeformed}	m\ddot{x}^{\mu}&=\mathcal{F}^{\mu}_{~~\nu}\dot{x}^{\nu}\nonumber \\
	&+\frac{2km}{3}\left[A\mathcal{F}^{\mu}_{~~\nu,\alpha}\dot{x}^{\nu}\dot{x}^{\alpha}+B\left(\mathcal{F}^{\mu}_{~~\nu}\mathcal{F}^{\nu}_{~~\alpha}\dot{x}^{\alpha}-\mathcal{F}^{\alpha}_{~~\nu}\mathcal{F}^{\nu}_{~~\beta}\dot{x}^{\beta}\dot{x}_{\alpha}\dot{x}^{\mu}\right)+\frac{2k}{3}C\mathcal{F}^{\mu}_{~~\nu,\alpha\sigma}\dot{x}^{\nu}\dot{x}^{\alpha}\dot{x}^{\sigma}\right]+{O}(k^4),
\end{align} 
where $A$, $B$ and $C$ are three parameters of dimensions $q_1^2/m_1^2$, $q_1^2/m_1^3$ and $q_1^4/m_1^3$, respectively. Note that the proposed deformation of the dynamics has two distinct aspects: (i) the deformation of the \textit{conservative} force field (or equivalently the potential); $\mathcal{F}^{(c)}_{\quad\mu\nu}\rightarrow\mathcal{F}_{\mu\nu}$ (or equivalently $\mathcal{A}^{(c)}_{\quad\mu}\rightarrow\mathcal{A}_{\mu}$) and (ii) the deformation of the \textit{radiation-reaction} force; accomplished by replacement of the coefficients $q_1^2/m_1^2$ $q_1^2/m_1^3$ and $q_1^4/m_1^3$ in the leading the radiation-reaction term in square brackets in the right-hand side of \ref{ADLmodified} by, respectively, $A$, $B$ and $C$. While the former captures the conservative part of the 2-body dynamics, the latter encodes the radiative effects. Like in gravity, we consider a static and radial external force field. Further, we can assume the following form for the deformed potential:
\begin{align}
	\mathcal{A}_{\mu}&=(\phi,0,0,0),\\\label{defphi}
	\phi(r)&=E\sum_{n=1}^{\infty}\left(\frac{k q_1q_2}{r M}\right)^n\phi_{n}(\nu,\gamma).
\end{align}
Note that in the limit $\nu\rightarrow 0$, the scalar potential $\phi$ must reduce to $kq_1q_2/r$. We shall explicitly see that this is indeed the case. The parameters $A$ and $B$ can be assumed in the following form:
\begin{align}\label{expandA}
	A&=\alpha _1(\nu,\gamma) \left(\frac{q_1^2}{m_1^2}+\frac{q_2^2}{m_2^2}\right)+\alpha_2(\nu,\gamma)\frac{ q_1 q_2}{m_1 m_2},\\\label{expandB}
	B&=\beta _1(\nu,\gamma) \left(\frac{q_1^2}{m_1^2m}+\frac{q_2^2}{m_2^2m}\right)+\beta_2(\nu,\gamma) \left(\frac{q_1^2}{m_1^3}+\frac{q_2^2}{m_2^3}\right)+\beta_3(\nu,\gamma)\frac{q_1 q_2}{m m_1 m_2}.
\end{align} 
Here, we expect $\alpha_1(0,\gamma)=\beta_1(0,\gamma)+\beta_2(0,\gamma)=1$, so that the Lorentz-Dirac equation is retained in the test-particle limit. 
To accommodate all potential terms of the required dimension that can be generated from $q_1,q _2,m_1,m_2$, one would have naively assumed an infinite series expansion for both $A$ and $B$ based on the dimension analysis. For instance, one could have added terms proportional to $q_1^3/(q_2m_2^2)$, $q_1^{100}/(q_2^{98}m_2^2)$, etc., to the ansatz for $A$. However, from the form of dissipative scattering observables $\delta J$ and $\delta E$, given by \ref{delJ} and \ref{delE}, respectively, one can guess that the finite expansions in \ref{expandA} and \ref{expandB} would suffice. On the other hand, the form of $C$ can only be fixed when the dissipative variables are known to $\mathcal{O}(k^4)$, which is currently unavailable. However, fortunately the parameter $C$ does not contribute to either $\chi$ or $\delta E$, up to $\mathcal{O}(k^3)$. Hence, we shall ignore this term, for the most part, barring some comments at a few junctures.

In summary, up to III-order, we have three sets of unknown dimensionless parameters in the EOB dynamics ---  $\{\phi_{1},\phi_{2},\phi_{3}\}$, $\{\alpha_1,\alpha_2\}$ and $\{\beta_1,\beta_2,\beta_3\}$. The next step is to determine the explicit forms of these parameters by matching the observables of the EOB problem to those of the original 2-body problem, as discussed in \ref{EOBkin}. 

\section{Effective-one-body formalism: the observables}\label{EOB_conclude}

The effective potential $\phi(r)$, as defined in \ref{defphi}, is characterized by an infinite set of coefficients $\{\phi_n| n=1,2,3,...\}$. However, as mentioned earlier, for results up to 3PL, the subset $\{\phi_1,\phi_2,\phi_3\}$ would suffice. We can fix $\{\phi_1,\phi_2,\phi_3\}$
by matching the conservative part of the scattering angle of the EOB problem to that of the original 2-body problem. While to fix $A$ and $B$, we shall do a similar exercise with the leading order radiated angular momentum and radiated energy. Therefore, our next step is to solve the scattering problem of the EOB system to 3PL. 

\subsection{Perturbative approach to the EOB dynamics}

As described in \ref{ADLdeformed}, we now solve the hyperbolic-like orbits of the reduced mass, interacting with an effective conservative force $\mathcal{F}(r)$ and a radiation-reaction force. Due to the system's symmetry, the entire scattering orbit will be constrained on a plane, where we introduce polar coordinate system $(r, \theta)$. As is standard in the study of orbital dynamics, for convenience, we introduce the variable $u\equiv 1/|\vec{x}|\equiv1/r$. Now, we want to obtain the function $u(\theta)$ that mathematically represents the orbit. In the spirit of the PL scheme, we shall solve the scattering orbit $u(\theta)$ as a series in the coupling constant. To this end, we expand as follows:
\begin{align}\label{expandu}
u(\theta)&=\frac{m\gamma}{j}\sin\,\theta+\left(\frac{kq_1q_2}{j}\right)\tilde{u}_1(\theta)+\left(\frac{kq_1q_2}{j}\right)^2\tilde{u}_2(\theta)+\left(\frac{kq_1q_2}{j}\right)^3\tilde{u}_3(\theta)+... \nonumber \\
&\equiv u_{0}(\theta)+k u_1(\theta)+k^2u_2(\theta)+k^3u_3(\theta)+...
\end{align}
where $j$ is the \textit{initial} angular momentum. Let us denote the angular momentum along the orbit of the EOB particle by $\mathcal{J}(\theta)$. Due to radiation reaction, $\mathcal{J}(\theta)$ is not a constant along the orbit. It is also convenient to expand the varying angular momentum $\mathcal{J}(\theta)$ in powers of $k$ as follows:
\begin{align}
    \mathcal{J}(\theta)\equiv m r^2\dot{\theta} &=j+\left(\frac{k q_1 q_2}{j}\right)^2\tilde{\mathcal{J}}_2(\theta)+\left(\frac{k q_1 q_2}{j}\right)^3\tilde{\mathcal{J}}_3(\theta)+ \cdots \nonumber \\
    &\equiv j+k^2\mathcal{J}_2(\theta)+k^3\mathcal{J}_3(\theta)+ \cdots 
\end{align}
Note that $\theta$-dependent corrections start only at $\mathcal{O}(k^2)$, the reason for which will be clear shortly. Similarly, we can also expand the energy function $\mathcal{E}(\theta)$ as:
\begin{align}
    \mathcal{E}(\theta)\equiv \varepsilon+\phi&=\epsilon+\left(\frac{k q_1 q_2}{j}\right)^2\tilde{\mathcal{E}}_2(\theta)+\left(\frac{k q_1 q_2}{j}\right)^3\tilde{\mathcal{E}}_3(\theta)+ \cdots \nonumber \\
    &\equiv \epsilon+k^2\mathcal{E}_2(\theta)+k^3\mathcal{E}_3(\theta)+ \cdots
\end{align}
where, $\varepsilon\equiv m\dot{t}$.  The evolution equations of $u(\theta)$, $\mathcal{J}(\theta)$ and $\mathcal{E}$ can be derived from \ref{ADLdeformed} and the mass-shell condition written in polar coordinates. The orbit is a solution to the following differential equation:
\begin{align}\label{OrbEOM}
    u''+u-\frac{\varepsilon \varepsilon'}{\mathcal{J}^2u'}+\frac{\varepsilon^2\mathcal{J}'}{\mathcal{J}^3u'}=0
\end{align}
where the prime ($'$) denotes derivative with respect to $\theta$. The rates of change of angular momentum and energy function take the forms:
\begin{align}\label{JEOM}
    \mathcal{J}'(\theta)&=\frac{2km}{3}\left[A\left(\frac{\varepsilon\mathcal{F} }{m^2u}\right)-B\left(\frac{\mathcal{F}^2}{u^2}+\frac{\mathcal{F}^2\mathcal{J}^2}{m^2}\right)+\frac{2k}{3}C\left(\frac{\mathcal{J}^2u'u^3\mathcal{F}}{m^2}+\frac{\mathcal{J}^2u'u^4\partial_{u}\mathcal{F}}{m^2}\right)\right]\\\label{EEOM}
    \mathcal{E}'(\theta)&=\frac{2km}{3}\left[A\left(\frac{\mathcal{J}u\mathcal{F}}{m}-\frac{\mathcal{J}(u')^2\partial_{u}\mathcal{F}}{m}\right)-B\left(\frac{\mathcal{J}\varepsilon \mathcal{F}^2}{m^2}\right)\right.\\\nonumber
    &\left.+\frac{2k}{3m^2}C\left(\mathcal{J}^2uu'\left(3 u\mathcal{F}+u\partial_{u}\mathcal{F})-(u')^2 (2 \partial_{u}\mathcal{F}+\partial^2_{u}\mathcal{F} u)\right)\right)\right]
\end{align}
Now, we can perturbatively solve the system of three equations \ref{OrbEOM}, \ref{JEOM}, and \ref{EEOM} in powers of $k$. To this end, we need to substitute \ref{expandu} into the three equations and iteratively solve the system. In particular, the equation of motion of the orbit at the $n-$th order can be cast in the following form:
\begin{align}
    u_{n}''+u_{n}+\mathcal{U}_{n}(u_{n-1},u_{n-2},\cdots,u_0)=0.
\end{align}
On the other hand, the equation describing the dissipation of energy and angular momentum simplifies to:
\begin{align}
    \mathcal{J}'_{n}&=\mathcal{T}_{n}(u_{n-2},u_{n-3},\cdots,u_0)\\
    \mathcal{E}'_{n}&=\mathcal{P}_{n}(u_{n-2},u_{n-3},\cdots,u_0)
\end{align}

\subsubsection{Comments on the \texorpdfstring{$C$}{C}-dependent RR term}\label{comment_on_C_RR}

Although we have displayed the $C$-dependent RR term in \ref{ADLdeformed} for generality, as we show in this subsection, the observables at $\mathcal{O}(k^3)$ are independent of $C$. To see this, 
we first rewrite the $C$-dependent RR term as a total derivative, modulo $\mathcal{O}(k^4)$ terms:
\begin{align}
 \frac{4k^2mC}{9}\mathcal{F}^{\mu}_{~~\nu,\alpha\sigma}\dot{x}^{\nu}\dot{x}^{\alpha}\dot{x}^{\sigma}=\frac{4k^2mC}{9}\frac{d}{d\tau}\left[\mathcal{F}^{\mu}_{~~\nu,\alpha}\dot{x}^{\nu}\dot{x}^{\alpha}\right]+\mathcal{O}(k^4)
\end{align}
Substituting the RHS of the above expression in  \ref{ADLdeformed}, we see that the contribution to the energy radiated from this term goes at most as $\lim_{\tau\rightarrow\infty}[u^3(\tau)-u^{3}(-\tau)]$, and hence, vanishes at $\mathcal{O}(k^3)$. Similarly, the $C-$dependent piece of the rate of change of angular momentum \ref{JEOM},
\begin{align}
    \frac{4k^2C}{9}\left(\frac{\mathcal{J}^2u'u^3\mathcal{F}}{m^2}+\frac{\mathcal{J}^2u'u^4\partial_{u}\mathcal{F}}{m^2}\right)\propto k^3\frac{d}{d\theta}u^6+\mathcal{O}(k^4) \, ,
\end{align}
is also a total derivative. Since $u\rightarrow 0$ at the asymptotes, it follows that the contribution of the above term to $\delta\mathcal{J}$ vanishes at $\mathcal{O}(k^3)$. Thus from the above two observations, one can naively conclude that $C$ cannot be fixed with the knowledge of $\mathcal{O}(k^3)$ observables. On the other hand, the same observations also indicate that we may predict the sub-leading contribution to $\delta J$ from only knowing the explicit forms of the coefficients $A$ and $B$, whose explicit forms are not available in the literature. In \ref{del_J3_predict}, we discuss this in detail.

\subsection{Scattering observables: the conservative parts}

We define the conservative scattering angle $\chi_{\rm cons}$ as the part of the total scattering angle $\chi$ that can be attributed to the dynamics generated by the conservative potential while neglecting the radiation reaction terms. The standard formula below gives the conservative scattering angle to be: 
\begin{align}\label{chiformula}
    \chi_{\rm cons}=\pi-2j\int_{r_{\rm min(c)}}^{\infty}\frac{dr}{r\sqrt{\left[(\epsilon-\phi)^2-m^2\right]r^2-j^2}}
\end{align}
where $r_{\min(c)}$ is the conservative piece of the value of radial coordinate at the closest approach and is given by:
\begin{align}
    r_{\min(c)}&=|b|+\frac{h_{\nu} q_1 q_2 \phi _1}{\gamma  m v^2}k+\frac{h_{\nu} q_1^2 q_2^2 \left(h_{\nu}
   \phi _1^2+2 \gamma ^2 \nu  v^2 \phi _2\right)}{2 |b| \gamma ^4 m^2
   v^4}k^2+\frac{h_{\nu}\nu  q_1^3 q_2^3 \left(\gamma  \nu  \phi _3-h_{\nu} \phi _1 \phi
   _2\right)}{|b|^2 \gamma ^2 m^3 v^2}k^3+\mathcal{O}(k^4)\\
   &\equiv |b|+b_1k+b_2k^3+b_3k^3+\mathcal{O}(k^4)
\end{align}
where, $b_1,b_2,b_3,..$ are constant coefficients defined by the above series.
A straightforward computation yields:
\begin{align}
    \chi_{\rm cons}&=\frac{2 k q_1 q_2 \phi_1 h_{\nu }}{|b|\gamma  m v^2}+\frac{\pi  k^2
   q_1^2 q_2^2 h_{\nu } \left(2 \gamma  \nu  \phi _2-h_{\nu} \phi _1^2\right)}{2
   |b|^2 \gamma ^2 m^2 v^2}+\\\nonumber
   &+\frac{2k^3q_1^3 q_2^3h_{\nu}}{3|b|^3\gamma ^6 m^3 v^6}\left[\phi_1h_{\nu}\left(\gamma\left(2\gamma ^2-3\right)\left(\phi_1^2h_{\nu}-6\gamma\nu\phi_2\right)-6 \nu\phi_2\right)+6\gamma^5\nu^2v^4\phi_3\right]+\mathcal{O}(k^4)
\end{align}
\subsection{Scattering observables: the dissipative parts}

From \ref{JEOM} and \ref{EEOM}, and noting that $\mathcal{F}$ is $\mathcal{O}(k)$, we see that $\mathcal{E}'$ and $\mathcal{J}'$ are $\mathcal{O}(k^2)$. Moreover, to solve $\mathcal{E}(\theta)$ and $\mathcal{J}(\theta)$ to $\mathcal{O}(k^3)$, we require only the orbital equation $u(\theta)$ to $\mathcal{O}(k)$. Solving \ref{OrbEOM} iteratively, we get:
\begin{align}
u(\theta)=\frac{m v\gamma}{j}\sin\,\theta+\left(\frac{k q_1 q_2}{j}\right)\left(\frac{h_{\nu}\phi_1 m\gamma}{j}\right)\left(\cos\,\theta-1\right)    +\mathcal{O}(k^2) \, .
\end{align}
Substituting the above in \ref{JEOM} and \ref{EEOM}, and integrating, we get:
\begin{align}
\mathcal{J}(\theta)&=j-\left(\frac{k q_1 q_2 }{j}\right)^2\left(\frac{2 j Am^2\gamma ^2 v \phi _1 h_{\nu } }{3q_1q_2 }\right)(\cos\theta -1)+\mathcal{O}(k^3)\\
\mathcal{E}(\theta)&=\epsilon-\left(\frac{k q_1 q_2 }{j}\right)^2\left(\frac{2 \gamma ^3 A m^3 v^3 \phi _1 h_{\nu } }{3 q_1q_2 }\right)\sin ^2\theta \cos\theta+\mathcal{O}(k^3) \, .
\end{align}
We have delegated the rather lengthy expressions for $\mathcal{J}_{3}$ and $\mathcal{E}_{3}$ to \ref{appendixa}. An interesting thing to note is that although the perturbations of both energy and angular momentum start at $\mathcal{O}(k^2)$,  the radiated angular momentum $\delta\mathcal{J}$ is $\mathcal{O}(k^2)$ while the radiated energy $\delta\mathcal{E}$ starts only at $\mathcal{O}(k^3)$. More precisely,
\begin{align}
    \delta\mathcal{J}&\equiv-[ \mathcal{J}(\pi-\chi)-\mathcal{J}(0)]\\\label{deltaJ3}
    &= -\left(\frac{k q_1 q_2}{j}\right)^2\frac{4 A j\gamma ^2 m^2  v \phi _1 h_{\nu }}{3 q_1 q_2 }\\\nonumber
&+\left(\frac{k q_1q_2}{j}\right)^3\frac{\pi  \gamma ^2 j m^2 \left[\phi _1^2 h_{\nu } \left(4 A \left(v^2+2\right)+B \left(3 \gamma ^2+1\right) m v^2\right)-8 A
   \gamma  \nu  v^2 \phi _2\right]}{12 q_1 q_2}+\mathcal{O}(k^4)\\
    \delta\mathcal{E}&\equiv-[ \mathcal{E}(\pi-\chi)-\mathcal{E}(0)]\\
    &=\left(\frac{k q_1 q_2}{j}\right)^3\frac{\pi  \gamma ^3 k^3 m^3 v^2 h_{\nu }^2 }{12 q_1^2q_2^2}\left(4 A+3 B \gamma ^2 m v^2\right)+\mathcal{O}(k^4) \, .
\end{align}
As mentioned in \ref{comment_on_C_RR}, the third order contribution to $\delta J$ is indeed independent of $C$. 

\section{Determining the EOB parameters}\label{conclude_EOB}

In the previous section, we perturbatively obtained the expressions for $u(\theta),  {\cal E}(\theta)$ and ${\cal J}(\theta)$, including the radiation-reaction terms. In this section, we determine the hitherto unknown co-coefficients $\{\phi_i\}, \{\alpha_i\}$, and $\{\beta_i\}$ in the EOB formulation. We also discuss how the EOB formalism can provide a way to obtain sub-leading contribution to the loss in angular momentum. 

To determine the first set of parameters $\{\phi_1,\phi_2,\phi_3\}$, we equate the conservative piece of the scattering angle of the original problem to that of the EOB one. This yields:
\begin{align}\label{phi1_found}
\phi_1&=1\\\label{phi2_found}
\phi_2&=\frac{h_{\nu }-1}{2 \gamma  \nu }\\\label{phi3_found}
\phi_3&=\frac{\left(2 \gamma
   ^4-3 \gamma ^2+1\right) \left(h_{\nu }-1\right) h_{\nu }-2
   \gamma ^5 \nu  v^4}{2 \gamma ^6 \nu ^2 v^4}
\end{align}
Similarly, we obtain the radiative coefficients $\{\alpha_i\}$ and $\{\beta_i\}$ using the matching conditions derived from the EOM mapping given in \ref{EOBkin}:
\begin{align}\label{EOBmap_deltaE}
   \frac{\delta\mathcal{E}}{h_{\nu}}  &= \delta E+\frac{(\delta E)^2}{2E}=\delta E+\mathcal{O}(k^4)\\\label{EOBmap_deltaJ}
    \frac{\delta\mathcal{J}}{h_{\nu}}&=\delta J+\frac{J}{E}\delta E+\frac{\delta E\delta J}{E}=\delta J+\mathcal{O}(k^3) \, .
\end{align}
The above mapping yields:
\begin{align}
\alpha_1&=\frac{1}{h_{\nu }}\quad;\quad \alpha_2=\frac{3 }{\gamma ^3 v^3 h_{\nu }}\left(\tanh ^{-1}(v)-\gamma ^2 v\right)\\
\beta_1&=\frac{3 \gamma ^2-4 \gamma  h_{\nu }+1}{3 \gamma ^3 v^2 h_{\nu }^2}\quad;\quad \beta_2=\frac{(\gamma -1) \left(3 \gamma ^2+1\right)}{3 \gamma ^3 v^2
   h_{\nu }^2}\\\nonumber
  \beta_3&= \frac{4 h_{\nu } \left[\gamma ^3 v-\gamma  \tanh ^{-1}(v)\right]+\left(3 \gamma ^2+1\right) \tanh ^{-1}(v)+\gamma v  [
   (4-3 \gamma ) \gamma^2 -9\gamma+4]}{\gamma ^6 v^5 h_{\nu }^2} \, .
\end{align}
This concludes our EOB formalism for the electromagnetic scattering problem, which accounts for the leading order radiative effects. 
\ref{app:Expansion} contains the expansion of the EOB potential in the test-particle limit. 

\subsection{Determining sub-leading radiated angular momentum} \label{del_J3_predict}

We now argue how the EOB formalism can be used to derive the sub-leading contribution to the angular momentum loss. As mentioned earlier, the direct computation of this $\mathcal{O}(k^3)$ term to $\delta J$ is currently unavailable. However, once the explicit forms of $A$ and $B$ are known, one can use \ref{deltaJ3} along with the EOB mapping from $\delta\mathcal{J}\rightarrow\delta J$ to `predict' the third-order contribution to $\delta J$. We find that this contribution $\delta J_{3}$ is:
\begin{align}\label{delta_J_3}
\delta J_{3}&=\left[\frac{J}{3m} \left(h_{\nu } \left(\frac{3}{\gamma }+\frac{4}{\gamma ^3 v^2}\right)-3 \nu \right) \right]\delta E_3+\left[\frac{\pi 
    q_1 q_2 \left(h_{\nu } \left(3 \gamma ^2 \left(v^2-2\right)+4\right)-3 \gamma ^2 v^2\right)}{12 \gamma ^2 J v h_{\nu
   }}\right]\delta J_2
\end{align}
where $\delta J_2$ and $\delta E_3$ correspond to the leading order radiated angular momentum \eqref{delJ} and energy \eqref{delE}.

\section{Results for bound orbits}\label{unbound_to_bound}

In the gravitational case, one of the most important applications of classical scattering results has been accurately forecasting bound-orbit dynamics. Several promising methods have been proposed to achieve this, including EFT methods~\cite{Neill:2013wsa,Amplitude2} and analytic continuations~\cite{Amplitude7,Kalin:2019inp}. Here, we explore using EOB formalism to extract binary-orbit results from the unbound-orbit ones. 

For the EM case, which is the focus of this work, binary orbits exist for $q_1q_2<0$. These binary orbits satisfy $E<M$, which in the EOB language translates to $\epsilon<m$.  The method initiated in Ref.~\cite{Amplitude7,Kalin:2019inp} corresponds to obtaining certain bound-orbit results by the analytical continuation of appropriate scattering observables into the domain $\gamma<1$. 
As a consistency check of our formalism, we now verify whether we can reproduce the bound orbits' results obtained using analytical continuation methods. In this section, we explicitly evaluate (1) the periastron shift $\Delta\Phi$ and (2) the energy radiated per orbital period ($\Delta E$), and (3) the angular momentum loss per orbital period ($\Delta J$) using the EOB formalism and compare them against the expressions evaluated using analytic continuation. For completeness, we have exhibited the expressions from the analytical continuation method in \ref{ub_to_b}. 

To compare bound-orbit results, it is convenient to employ certain useful variables. We start with the mass-subtracted energy (or `the non-relativistic energy') $E_{nr}\equiv E-M$. The bound orbits correspond to those with $E_{nr}<0$. Recalling that $E_{nr}<0$ corresponds to $\epsilon/m=\gamma<1$, the function $h_{\nu}=E/M=\sqrt{1+2\nu(\gamma-1)}$ for the bound orbits satisfies $h_{\nu}<1$. Since $h_{\nu}$ and $\gamma$ were originally defined for the scattering problem (for which $h_{\nu},\gamma\geq 1$), we introduce a subscript $(b)$ for these variables whenever they specifically refer to bound orbits; i.e., $h_{\nu(b)}$ and $\gamma_{(b)}$ for $E_{nr}<0$. Moreover, for bound orbits $v=iv_{b}$, where $v_{b}\in\mathbb{R}$. 

\subsection{Periastron shift}

The leading order periastron shift is expected at $\mathcal{O}(k^2)$. So, we shall next find the \textit{conservative} bound-orbit at this order. The (conservative) equation of motion for the orbit takes the form:
\begin{align}
    u''&+\omega^2u+\left(\frac{k q_1q_2}{j^2}\right)\gamma_{(b)}m \phi _1 h_{\nu(b)}+\mathcal{O}(k^3)=0,\\
    \omega^2&=1-\left(\frac{k q_1q_2}{j^2}\right)^2 h_{\nu(b) } \left(\phi _1^2 h_{\nu(b)}-2 \gamma_{(b)}  \nu  \phi _2\right) ,
\end{align}
where the subscript`$(b)$' with $\gamma$ and $h_{\nu}$ refers to bound orbits (i.e., $\gamma_{(b)},h_{\nu(b)}<1$). The orbital equation is:
\begin{align}\label{bound_orbit}
    u_{b}(\theta)=\bar{u} \cos[\omega(\theta-\bar{\theta})]-\frac{\gamma_{(b)}  k m q_1 q_2 \phi _1 h_{\nu(b) }}{j^2}
\end{align}
where the constant $\bar{u}$ can be found by demanding the mass-shell condition and $\bar{\theta}$ just captures the initial condition. Therefore, the periastron shift at $\mathcal{O}(k^2)$ can be found from:
\begin{align}
    \Delta\Phi&=\frac{2\pi}{\omega}-2\pi+\mathcal{O}(k^4)\\
    &=\frac{\pi  k^2 q_1^2 q_2^2 h_{\nu(b) } \left(\phi _1^2 h_{\nu(b)}-2 \gamma_{(b)}  \nu  \phi _2\right)}{j^2}+\mathcal{O}(k^4)\\
    &=\frac{\pi k^2q_1^2q_2^2}{J^2h_{\nu(b)}}+\mathcal{O}(k^4)
\end{align}
where, in the last line we have used \ref{phi1_found} and \ref{phi2_found}, along with the EOB map $J=h_{\nu} j$.  Note that the above expression matches exactly with \ref{per_shift_continuation}. 

\subsection{Energy loss per orbit}

\ref{EEOM} gives the general expression for the rate of change of energy. Hence, the total energy \textit{loss} per orbital period of the EOB particle $\Delta\mathcal{E}$ is:
\begin{align}
    \Delta \mathcal{E}&=-\oint \mathcal{E}'d\theta \, .
\end{align}
Substituting the bound-orbit solution $u_{b}(\theta)$ from \ref{bound_orbit} into the above equation yields:
\begin{align}
   \Delta \mathcal{E}= \frac{\pi  \gamma_{(b)}  \left(\gamma_{(b)} ^2-1\right) k^3 m^3 q_1^2 q_2^2 \phi _1^2 h_{\nu(b) }^2 \left(4 A+3 B \left(\gamma_{(b)} ^2-1\right) m\right)}{6 j^3} +\mathcal{O}(k^4) \, .
\end{align}
Using the EOB mapping in \ref{EOBmap_deltaE}, the above equation translates to:
\begin{align}
    \Delta E&=\frac{\pi m^3k^3q_1^2q_2^2(\gamma_{(b)}^2-1)}{2 J^3h^4_{\nu(b)}}\left[\frac{(3\gamma_{(b)}^2+1)}{3}\left(\frac{q_1^2}{m_1^2}+\frac{q_2^2}{m_2^2}\right)\right.\\\nonumber
    &\left.+\frac{(\gamma_{(b)}-1)(3\gamma_{(b)}^2+1)}{3}\left(\frac{q_1^2m}{m_1^3}+\frac{q_2^2m}{m_2^3}\right)-\mathcal{G}(\gamma_{(b)})\frac{q_1q_2}{m_1m_2}\right]+\mathcal{O}(k^4)
\end{align}
where we have also used the explicit forms of $A$ and $B$ as found in \ref{conclude_EOB} to arrive at the final expression. The above expression matches exactly with that in \ref{energy_loss_continuation}.

As a further consistency check, it is instructive to consider the non-relativistic limit of the above expression. To this end, recall the definition of non-relativistic energy $E_{nr}\equiv E-M$, so that $h_{\nu(b)}=E_{nr}/M+1$. Therefore, the non-relativistic limit $\Delta E_{nr}$ of the energy loss takes the form:
\begin{align}
    \Delta E&= \Delta E_{nr}+\mathcal{O}\left(\frac{E_{nr}^2}{M^2}\right),\\
    \Delta E_{nr}&=\frac{4\pi m^2k^3q_1^2q_2^2E_{nr}}{3J^3}\left(\frac{q_1}{m_1}-\frac{q_2}{m_2}\right)^2+\mathcal{O}(k^4) \, .
\end{align}
The above expression matches perfectly with the leading order non-relativistic limit of energy loss computed in \ref{NR_loss}.   

\subsection{Angular momentum loss per orbit}
\ref{JEOM} describes the rate of change of angular momentum. The angular momentum lost per orbit of the EOB particle is:
\begin{align}
    \Delta \mathcal{J} =-\oint \mathcal{J}'d\theta \, .
\end{align}
Substituting the bound orbit $u_{b}(\theta)$ from \ref{bound_orbit} into the above equation, we get:
\begin{align}
    \Delta \mathcal{J}&=\frac{2 \pi  A k^3 m^2 q_1^2 q_2^2 h_{\nu(b) } \left(\gamma_{(b)} ^2+2 \gamma_{(b)} ^2 h_{\nu(b) }-1\right)}{3 j^2}+\frac{\pi  B \left(3 \gamma ^4-2 \gamma_{(b)}
   ^2-1\right) k^3 m^3 q_1^2 q_2^2 h_{\nu(b) }^2}{6 j^2}+\mathcal{O}(k^4) \, .
\end{align}
Now, using the EOB mapping in \ref{EOBmap_deltaJ}, we can evaluate the angular momentum loss per orbit of the original binary system:
\begin{align}
\Delta J&=\frac{2 \pi  A k^3 m^2 q_1^2 q_2^2 h_{\nu(b) } \left(-\nu\gamma_{(b)} ^3+\gamma_{(b)}  \nu +h_{\nu(b) } \left(\gamma_{(b)} ^2+2 \gamma_{(b)} ^2 h_{\nu(b)
   }-1\right)\right)}{3 J^2}\\\nonumber
   &-\frac{\pi  B \left(\gamma_{(b)} ^2-1\right) k^3 m^3 q_1^2 q_2^2 h_{\nu(b) } \left(3 \gamma_{(b)}  \left(\gamma_{(b)} ^2-1\right) \nu
   -\left(3 \gamma_{(b)} ^2+1\right) h_{\nu(b) }^2\right)}{6 J^2} +\mathcal{O}(k^4) \, .
\end{align}
Substitution of $A$ and $B$ in the above equation produces a lengthy expression, which is not illuminating and will not be given here. However, the final expression thus obtained matches exactly the one obtained from \ref{angmom_loss_continuation}. In other words, $\Delta J$ from our EOB formalism is precisely $2k^3\delta J_3+\mathcal{O}(k^4)$. 

As in the case of energy loss, the non-relativistic limit of $\Delta J$ yields:
\begin{align}
    \Delta J&=\Delta J_{nr}+\mathcal{O}\left(\frac{E_{nr}}{M}\right)\\
    \Delta J_{nr}&=\frac{4\pi k^3 m^2q_1^2q_2^2}{3J^2}\left(\frac{q_1}{m_1}-\frac{q_2}{m_2}\right)^2 \, .
\end{align}
The above expression for $\Delta J_{nr}$ matches precisely with the explicit non-relativistic computation of angular momentum loss given in \ref{NR_loss}.

\section{Comparison to other approaches}\label{review_comparison}

In the previous section, we demonstrated how to use the EOB formalism to derive physical quantities for the binary-orbit from the unbound-orbit ones. We also showed that the EOB formalism gives identical results to the other approaches. In this section, we compare and contrast our approach to certain other ones, although mainly concerning gravitational dynamics, already available in the literature. 

\subsection{Effective-field-theory inspired approach}\label{EFT_comparison}

The effective-field-theory (EFT) based approach for gravitational binaries of non-rotating compact objects was pioneered by Goldberger, and Rothstein \cite{PhysRevD.73.104029,Goldberger:2006bd}. The extension of the approach that accounts for the spin was also later developed\cite{PhysRevD.73.104031,porto2007effective}. For a modern review of the EFT approach to gravitational dynamics, consult \cite{Porto:2016pyg}. EFT methods have recently been used to study PM dynamics\cite{Kalin:2020mvi,Kalin:2020fhe,Kalin:2020lmz,Liu:2021zxr,Dlapa:2021npj,Dlapa:2021vgp}, even including the radiation effects\cite{Kalin:2022hph,Jakobsen:2022psy,Dlapa:2022lmu,Jakobsen:2021smu,Jakobsen:2022fcj,Mougiakakos:2021ckm,Riva:2021vnj,Riva:2022fru}. 

In \ref{EOBdyn}, we invoked \ref{ADLdeformed} as a deformation of the standard Lorentz-Dirac equation to account for the recoil effects. However, one can also motivate the same deformed Lorentz-Dirac equation from EFT-inspired reasoning. To this end, instead of referring to the Lorentz-Dirac equation, we can seek the most general form of the radiation-reaction force that is allowed by symmetries of the system. More specifically, we look for additions to the Lorentz force equation that are: (1) Lorentz-invariant, (2) gauge-invariant, (3) orthogonal to $\dot{x}^{\nu}$. These assumptions imply that the allowed terms, up to $\mathcal{O}(k^3)$, are proportional to:
\begin{align}
\left\{A \mathcal{F}^{\mu}_{~~\nu,\alpha}\dot{x}^{\nu}\dot{x}^{\alpha},B\left(\mathcal{F}^{\mu}_{~~\nu}\mathcal{F}^{\nu}_{~~\alpha}\dot{x}^{\alpha}-\mathcal{F}^{\alpha}_{~~\nu}\mathcal{F}^{\nu}_{~~\beta}\dot{x}^{\beta}\dot{x}_{\alpha}\dot{x}^{\mu}\right),k C\mathcal{F}^{\mu}_{~~\nu,\alpha\sigma}\dot{x}^{\nu}\dot{x}^{\alpha}\dot{x}^{\sigma}\right\}
\end{align}
where, $A$, $B$ and $C$ are three parameters of dimensions $q_1^2/m_1^2$, $q_1^2/m_1^3$ and $q_1^4/m_1^3$, respectively. By extending this logic, one can write down the radiation reaction terms to any desired order and introduce appropriate parameters, which can be fixed by matching observables of the EOB problem to that of the original 2-body one. 

\subsection{Conventional approaches to EOB for dissipative dynamics }\label{conv_comparison}

The dissipative effects due to radiation can be accounted for by adding an appropriate radiation reaction term to the equation of motion. Retaining our notation for the polar coordinates for the relative coordinate, the radiation reaction $\mathbf{f}$ force can be generically written as $\mathbf{f}=f_r\hat{r}+(f_{\theta}/r)\hat{\theta}$. The rate of change of total angular momentum and energy takes the following form~\cite{Amplitude18,Bini:2012ji}:
\begin{align}
    \frac{d \mathcal{E}}{d t}&=\frac{dr}{d t}f_{r}+\frac{d \theta}{d t}f_{\theta}\\
    \frac{d \mathcal{J}}{d t}&=f_{\theta} \, .
\end{align}
Finding the components $\{f_{r},f_{\theta}\}$ of the radiation reaction force requires balancing the energy and angular momentum emitted to infinity by radiation with that dissipated by the EOB system.
In the gravity case, this PM-inspired approach was used to determine the radiation response force as a series in $G$. 
To facilitate this, in Ref.~\cite{Manohar:2022dea}, the RR force components were written in terms of two functions $c_{r}$ and $c_{p}$, such that:
\begin{align}\label{RR_conventional}
    \mathbf{f}\equiv (c_r+c_p)p_{r}\hat{r}+\frac{c_p \mathcal{J}}{r}\hat{\theta}
\end{align}
where $p_{r}$ is the radial component of the relative momentum. The idea is to find the functions $\{c_r,c_p\}$ by an appropriate matching of the energy and angular momentum radiated as gravitational waves to, respectively, the energy and angular momentum lost by the binary system. In the spirit of PM and further assuming $\{c_r,c_p\}$ to be functions of only $r$ and $|p|^2$, Ref.~\cite{Manohar:2022dea} describes the general procedure to find these functions as a series in $G$ and gives explicit results up to $\mathcal{O}(G^2)$. 

One can employ a similar approach for the EM case as well. As opposed to the top-down approach for RR force in Ref.~\cite{Manohar:2022dea}, our method begins with the most general expression for the RR force at the desired order, so the series expansions of the corresponding functions $\{c_r,c_p\}$ are straightforward and follow directly from expanding the appropriate components of \ref{ADLdeformed}. To this end, we first assume the functions $c_{r}$ and $c_{p}$ to have the following general expansion:
\begin{align}\label{cr_expansion}
c_{r}&=\frac{1}{r}\left[\left(\frac{k q_1 q_2}{M r}\right)^2c_{(2)r}+  \left(\frac{k q_1 q_2}{M r}\right)^3c_{(3)r}+ ...  \right] \\\label{cp_expansion}
c_{p}&=\frac{1}{r}\left[\left(\frac{k q_1 q_2}{M r}\right)^2c_{(3)p}+  \left(\frac{k q_1 q_2}{M r}\right)^3c_{(3)p}+ ...  \right] 
\end{align}
On the other hand, the radial and angular components of the RR force in \ref{ADLdeformed} which, ignoring the $C-$dependent term, take the forms:
\begin{align}
f_{r}&= \left[\frac{2kA}{3}\partial_{r}\mathcal{F}-\frac{2kB\mathcal{J}^2}{3mr^2\varepsilon}\mathcal{F}^2\right]p_{r}+\\
f_{\theta}&=\left[\frac{2kA\mathcal{F}}{3r}-\frac{2kmB\mathcal{F}^2}{\varepsilon}\left(1+\frac{\mathcal{J}^2}{m^2r^2}\right)\right]\mathcal{J}
\end{align}
where, $\mathcal{F}=-\partial_{r}\phi$ and $\varepsilon=m\dot{t}$. In the above equation, we can use the definition of $\{c_r,c_p\}$ from \ref{RR_conventional} and the series expansion for the $\phi$ obtained in \ref{conclude_EOB} to arrive at:
\begin{align}\label{cr_expression}
c_{r}&=\frac{1}{r}\left[-\frac{2 A k^2 q_1 q_2 \phi _1 h_{\nu }}{ r^2}+\left(-\frac{16 A k^3 \nu  q_1^2 q_2^2 \phi _2 h_{\nu }}{3 m r^3}+\frac{2 B k^3 q_1^2 q_2^2 \phi _1^2 h_{\nu }^2}{3 \gamma  r^3}\right)+\mathcal{O}(k^4)\right]\\\label{cp_expression}
    c_{p}&=\frac{1}{r}\left[\frac{2 A k^2 q_1 q_2 \phi _1 h_{\nu }}{3 r^3}+\left(\frac{4 A k^3 \nu  q_1^2 q_2^2 \phi _2 h_{\nu }}{3 m r^3}-\frac{2 B k^3 q_1^2 q_2^2 \phi _1^2 h_{\nu }^2 \left(j^2+m^2 r^2\right)}{3 \gamma  m^2 r^5}\right)+\mathcal{O}(k^4)\right]
\end{align}
where, $A$ and $B$ are also as found in \ref{conclude_EOB}. 

We can now compare our findings to those related to gravity~\cite{Manohar:2022dea}: First, in Ref. ~\cite{Manohar:2022dea}, it was found that demanding $\delta\mathcal{E}$ and $\delta\mathcal{J}$ at $\mathcal{O}(G^2)$ in the gravitational case fixes the RR force entirely at that order. For instance, the fact that $\delta\mathcal{E}$ vanishes at $\mathcal{O}(k^2)$ imply that $c_{p}=-3c_{r}+\mathcal{O}(G^4)$. Using \ref{cr_expression} and \ref{cp_expression}, it is easy to see that the analogous condition, namely, $c_{p}=-3c_{r}+\mathcal{O}(k^4)$ is automatically satisfied in our formalism.

\section{Summary and future outlook}\label{summary}

We have described the effective-one-body formalism for conservative and radiative dynamics of a relativistic electromagnetic binary system. As a concrete illustration of the formalism, we discussed the details of EOB dynamics at $\mathcal{O}(k^3)$. 
At this order, the system's symmetry requires that the EOB dynamics be parameterized by three conservative parameters  $\{\phi_1,\phi_2,\phi_3\}$ and three dissipative parameters $\{A, B, C\}$.
While the former set describes the conservative potential of the EOB dynamics, the latter captures the radiation-reaction force. However, physical arguments show that the parameter $C$ is irrelevant for observables at $\mathcal{O}(k^3)$. By matching the conservative part of the $\mathcal{O}(k^3)$ scattering angle of the original two-body problem and that of the EOB system, we found the explicit forms of the parameters $\{\phi_1,\phi_2,\phi_3\}$. Similarly, by comparing the $\mathcal{O}(k^3)$ radiated energy and $\mathcal{O}(k^2)$ angular momentum on both sides, following the EOB-mapping reviewed in \ref{EOBkin}, we found the explicit forms of the dissipative parameters $A, B$. With the exact values of $\{\phi_1,\phi_2,\phi_3\}$ and $A, B$, our formalism describes the full dynamics of an electromagnetically charged binary system at $\mathcal{O}(k^3)$, including certain non-perturbative aspects. As a further application and cross-check, we have studied the bound orbit dynamics of the system at $\mathcal{O}(k^3)$ using our formalism. To this end, we focused on calculating three observables: the periastron shift ($\Delta\phi$), radiated energy per orbit ($\Delta E$) and radiated angular momentum per orbit ($\Delta J$). Our results for these observables match perfectly with the ones expected from the method of unbound-to-bound analytical continuation. 

Interestingly, our formalism leads to a conjecture for the subleading contribution to the net angular momentum loss for the unbound orbits ($\delta J_3$) and the leading order angular momentum loss per orbit ($\Delta J$) for the bound orbits, whose explicit computations are unavailable yet. We verified that the non-relativistic limit of our expression for $\Delta J$ matches precisely with that obtained by explicit non-relativistic calculations, potentially indicating that our conjecture is accurate. As noted in Ref.~\cite{Saketh:2021sri}, the explicit computation of $\delta J_3$ (and, by analytical continuation, $\Delta J$) requires first solving the exact forms of the worldlines of the original scattering particles at $\mathcal{O}(k^3)$. The solving of the worldlines at 3PL involves rather cumbersome integrals. However, if correct, the EOB approach offers an extremely economical way to derive $\delta J_3$ and $\Delta J$.    

In \ref{EFT_comparison}, we have briefly discussed a plausible alternate interpretation of the EOB dynamics in terms of the EFT framework. However, it is desirable to formalize the reasoning therein, which will be the subject of a forthcoming publication. Here, we will briefly outline how one could achieve this objective. The standard Lagrangian-based EFT approach cannot efficiently account for dissipative effects. An elegant Lagrangian formulation of classical dissipative systems, inspired by the Schwinger-Keldysh formalism for non-equilibrium quantum systems, was proposed in \cite{PhysRevLett.110.174301}. For the EOB particle considered in this work, the application of this formalism starts with the doubling $x^{\mu}\rightarrow \{x^{\mu}_{1},x^{\mu}_{2}\}$, while the Lagrangian takes the form:
\begin{align}
    \mathcal{L}=\mathcal{L}_{EM}(x^{\mu}_1,\dot{x}^{\mu}_1)-\mathcal{L}_{EM}(x^{\mu}_1,\dot{x}^{\mu}_1)+K(x^{\mu}_i,\dot{x}^{\mu}_i)
\end{align}
where $i=1,2$, $\mathcal{L}_{Em}$ is the standard quadratic Lagrangian for a particle in an external electromagnetic field, and $K$ is a function that cannot be written as a difference of the form $f(x^{\mu}_1,\dot{x}^{\mu}_1)-f(x^{\mu}_2,\dot{x}^{\mu}_2)$. The radiation reaction terms can be accounted for by an appropriate choice of $K(x^{\mu}_i,\dot{x}^{\mu}_i)$. This choice, in turn, may be further constrained by the symmetries of the systems and expanded in powers of coupling $k$ in a manner analogous to the standard EFT approach. Presumably, such a systematic procedure would lead to an alternate, more rigorous justification for the deformed Lorentz-Dirac equation given in \ref{ADLdeformed}. Further, it would be interesting to see if the RR coefficients (i.e., $A$, $B$, $C$, etc.) can be directly related to the scattering amplitude analogous to the conservative potential extracted from amplitudes.

We are confident that the formalism outlined here for the EM case can be extended to gravity. However, one immediate hurdle to a naive extension of our approach to the GR case is that even in the test particle limit, the leading order RR force, as, for instance, described by the MiSaTaQuWa equation\cite{PhysRevD.55.3457,PhysRevD.56.3381}, is not a simple local function of the worldline of the particle. Rather, it depends on the history of the particle as well. Nevertheless, one might seek a suitable deformation of the MiSaTaQuWa equation that can serve as the RR term in the EOB equation of motion. 

Alternatively, one might employ the series expansion approach described in \cite{Manohar:2022dea} (and briefly revised in \ref{conv_comparison}). Regarding this, in Ref.~\cite{Manohar:2022dea}, it was assumed that $c_{r}$ and $c_{p}$ are functions of $r$ and $|\vec{p}|^2$. It appears that the EM analog of this assumption is inconsistent with the expressions for $c_{r}$ and $c_{p}$ given by \ref{cr_expression} and \ref{cp_expression}. It seems more reasonable to regard $\{c_{(i)r},c_{(i)p}\}$ as functions of both $|\vec{p}|^2$ and $(\mathcal{J}^2/m^2r^2)$. Consequently, the assumption that the RR force can be written in radial gauge appears to be not true, at least in the EM case. In GR, this assumption (of dependence on only $r$ and $|\vec{p}|^2$) may hold while the same is violated in EM. If this is the case, it merits additional investigation and will be the subject of subsequent studies. 
 

\section*{Acknowledgement}
The work is supported by SERB-MATRICS grant. 

\appendix	
\labelformat{section}{Appendix #1}
\labelformat{subsection}{Appendix #1}
\labelformat{subsubsection}{Appendix #1}

\renewcommand{\theequation}{A.\arabic{equation}}

\setcounter{equation}{0}
\section*{Appendices}

In the following sections, we outline some of the mathematical details concerning our work that are not explicitly discussed in the main text. 
\section{Third-order energy and angular momentum}\label{appendixa}
The third-order angular momentum is given by:
\begin{align}
    \mathcal{J}_{3}(\theta)&=A\mathcal{J}_{3(A)}(\theta)+B\mathcal{J}_{3(B)}(\theta)+C\mathcal{J}_{3(C)}(\theta)\\
    \mathcal{J}_{3(A)}&=\frac{\gamma ^2 m^2 q_1^2 q_2^2 h_{\nu } \left(\phi _1^2 h_{\nu } \left(4 \sin (\theta )-2 \theta  \left(v^2+2\right)+v^2 \sin (2 \theta  )\right)+2 \gamma  \nu  v^2 \phi _2 (2 \theta -\sin (2 \theta ))\right)}{6 j^2}\\
    \mathcal{J}_{3(B)}&=-\frac{\gamma ^2 m^3 q_1^2 q_2^2 v^2 \phi _1^2 h_{\nu }^2 \left(4 \theta  \left(3 \gamma ^2 v^2+4\right)+\gamma ^2 v^2 \sin (4 \theta )-8
   \sin (2 \theta ) \left(\gamma ^2 v^2+1\right)\right)}{48 j^2}\\
   \mathcal{J}_{3(C)}&=\frac{4 \gamma ^6 m^5 q_1 q_2 v^6 \phi _1 h_{\nu } \sin ^6(\theta )}{9 j^4}
\end{align}
The third-order energy turns out to be:
\begin{align}
    \mathcal{E}_{3}&=A\mathcal{E}_{3(A)}(\theta)+B\mathcal{E}_{3(B)}(\theta)+C\mathcal{E}_{3(C)}(\theta)\\
    \mathcal{E}_{3(A)}(\theta)&=\frac{\gamma ^3 h m^3 q_1^2 q_2^2 v^2 \left(h \phi _1^2 (-2 \theta +\sin (\theta )+5 \sin (2 \theta )-3 \sin (3 \theta ))-8 \gamma  \nu  v^2
   \phi _2 \sin ^3(\theta ) \cos (\theta )\right)}{6 j^3}\\
   \mathcal{E}_{3(B)}(\theta)&=-\frac{\gamma ^5 h^2 m^4 q_1^2 q_2^2 v^4 \phi _1^2 (12 \theta -8 \sin (2 \theta )+\sin (4 \theta ))}{48 j^3}\\
   \mathcal{E}_{3(C)}(\theta)&=-\frac{2 \gamma ^5 h m^4 q_1 q_2 v^5 \phi _1 \sin ^3(\theta ) (3 \cos (2 \theta )+1)}{9 j^3}
\end{align}
\renewcommand{\theequation}{B.\arabic{equation}}

\setcounter{equation}{0}

\section{Non-relativistic limit of angular momentum and energy loss}\label{NR_loss}
The dipole formula for the rate of radiated angular momentum loss is\cite{Landau:1975pou}:
\begin{align}
    \frac{d\mathbf{J}}{dt}&=-\frac{2k}{3}\mathbf{D}\times\dddot{\mathbf{D}}
\end{align}
where, $\mathbf{D}$ is the dipole moment vector. For a binary system of charges, we have:
\begin{align}
    \mathbf{D}=\left(\frac{q_1}{m_1}-\frac{q_2}{m_2}\right)m\mathbf{r}
\end{align}
where $\mathbf{r}$ is the position vector of particle 2 with respect to particle 1. In the non-relativistic limit, the force between the charges is the simple Coulomb force. Using this fact, the rate of angular momentum loss :  
\begin{align}
    \frac{dJ}{d\theta}&=-\frac{2k^2mq_1q_3}{3r}\left(\frac{q_1}{m_1}-\frac{q_2}{m_2}\right)^2
\end{align}
The non-relativistic orbit is given by $r=R/(1+\sigma \cos\theta)$, where $\sigma^2=1+\frac{2E_{nr}J^2}{m(kq_1q_2)^2}$ and $R=J^2/(k q_1q_2m)$. Integrating the above equation over one orbital period, we find that the non-relativistic limit of angular momentum loss per orbit is given by:
\begin{align}
    \Delta J_{nr}=\frac{4 \pi  k^3 m^2 q_1^2 q_2^2}{3 J^2}\left(\frac{q_1}{m_1}-\frac{q_2}{m_2}\right)^2
\end{align}

The dipole energy loss, on the other hand, has the 
 following expression\cite{Landau:1975pou}:
 \begin{align}
     \frac{d E}{dt}=-\frac{2k}{3}\dddot{\mathbf{D}}.\dot{\mathbf{D}}
 \end{align}
 Again, using the fact that the force between the particles is Coulombian, the above equation simplifies to:
 \begin{align}
     \frac{d E}{d\theta}=\frac{2 k^3 m q_1^2 q_2^2}{3 J r^2}\left(\frac{q_1}{m_1}-\frac{q_2}{m_2}\right)^2
 \end{align}
 Upon integrating the above equation for one orbit, we get:
 \begin{align}
     \Delta E_{nr}=\frac{4\pi m^2k^3q_1^2q_2^2E_{nr}}{3J^3}\left(\frac{q_1}{m_1}-\frac{q_2}{m_2}\right)^2+\mathcal{O}(k^4)
 \end{align}
\renewcommand{\theequation}{C.\arabic{equation}}

\setcounter{equation}{0}
\section{Expansion about the test-particle limit}
\label{app:Expansion}

It is instructive to look at expansions of the EOB parameters in the symmetric mass ratio $\nu$. The effective potential takes the form:
\begin{align}
    \phi(r)&=\frac{k q_1q_2}{r}\\\nonumber
    &+\nu\left[\frac{(\gamma -1) k q_1 q_2}{r}+\frac{(\gamma -1) k^2 q_1^2 q_2^2}{2\gamma  m r^2}-\frac{(2 \gamma +1) k^3 q_1^3 q_2^3}{2 \gamma ^2 (\gamma +1)m^2 r^3}+\mathcal{O}(k^4)\right]+\mathcal{O}(\nu^2),
\end{align}
which has the leading order term consistent with that expected from the test-particle limit. On the other hand, the radiative coefficients have the expansions:
\begin{align}
    \alpha_1&=1+(1-\gamma ) \nu +\mathcal{O}\left(\nu ^2\right)\\
    \alpha_2&=\frac{3 \left(\tanh ^{-1}(v)-v \gamma ^2\right)}{v^3 \gamma ^3}+\frac{3 (\gamma -1) \left(v
   \gamma ^2-\tanh ^{-1}(v)\right) \nu }{v^3 \gamma ^3}+\mathcal{O}\left(\nu ^2\right)\\
   \beta_1+\beta_2&=1+\frac{\left(2-6 \gamma ^2\right) \nu }{3 \gamma +3}+\mathcal{O}(\nu^2)\\
   \beta_3&=\frac{\nu  \left(\left(-6 \gamma ^2+4 \gamma -2\right) \cosh ^{-1}(\gamma )
   +2 \gamma  (3 \gamma  ((\gamma -2) \gamma +3)-4) v\right)}{\gamma
   ^4 (\gamma +1) v^3}\\\nonumber
   &+\frac{(3 \gamma -1) \cosh ^{-1}(\gamma )+\gamma  ((5-3 \gamma ) \gamma -4) v}{\gamma ^4 (\gamma +1) v^3}+\mathcal{O}(\nu^2)
\end{align}
In particular, at the leading order, $\alpha_1$ and $\beta_1+\beta_2$ are both unity, which is also consistent with the expected test-particle limit.

\renewcommand{\theequation}{D.\arabic{equation}}

\setcounter{equation}{0}

\section{Results for bound orbits from other methods}
\label{ub_to_b}

In this appendix, we give the results of the analyical continuation method for (1) the periastron shift $\Delta\Phi$~\cite{Amplitude7,Kalin:2019inp} and (2) the energy radiated per orbital period ($\Delta E$)~\cite{Bini:2020hmy,Saketh:2021sri}, and (3) the angular momentum loss per orbital period ($\Delta J$)~\cite{Saketh:2021sri}. Following Refs. \cite{Amplitude7,Kalin:2019inp}, the periastron shift can be found from the scattering angle by analytically continuing  $E_{nr}$, and  $J$:
\begin{align}
    \Delta\Phi(E_{nr},J)=-\chi(E_{nr},J)-\chi(E_{nr},-J).
\end{align}
Note that the left-hand-side is defined for $E_{nr}<0$, while the right-hand-side is written in terms of functions originally defined for $E_{nr}>0$, but analytically continued to the bound-orbit domain $E_{nr}<0$. Hence, from \ref{chi_cons}, the leading-order periastron shift takes the following form:
\begin{align}\label{per_shift_continuation}
 \Delta\Phi=\frac{\pi k^2q_1^2q_2^2}{J^2h_{\nu(b)}}+\mathcal{O}(k^4) \, .
\end{align}
Similarly, following Refs.~\cite{Bini:2020hmy,Saketh:2021sri}, the energy loss per orbit can be obtained by analytically continuing the energy, i. e., 
\begin{align}\label{energy_loss_continuation}
    \Delta E(E_{nr},J)=\delta E(E_{nr},J)-\delta E(E_{nr},-J).
\end{align}
As in the expression for periastron shift, on the left-hand-side is a function naturally defined for bound orbits ($E_{nr}<0$), while the right-hand-side is constructed from analytic continuations of functions originally defined for scattering orbits ($E_{nr}>0$). From \ref{delE}, 
the energy loss per orbit is then given by:
\begin{align}
    \Delta E&=\frac{\pi m^3k^3q_1^2q_2^2(\gamma_{(b)}^2-1)}{2 J^3h^4_{\nu(b)}}\left[\frac{(3\gamma_{(b)}^2+1)}{3}\left(\frac{q_1^2}{m_1^2}+\frac{q_2^2}{m_2^2}\right)\right.\\\nonumber
    &\left.+\frac{(\gamma_{(b)}-1)(3\gamma_{(b)}^2+1)}{3}\left(\frac{q_1^2m}{m_1^3}+\frac{q_2^2m}{m_2^3}\right)-\mathcal{G}(\gamma_{(b)})\frac{q_1q_2}{m_1m_2}\right]+\mathcal{O}(k^4)
\end{align}
Likewise, the loss of angular momentum per orbit is~\cite{Saketh:2021sri}:
\begin{align}\label{angmom_loss_continuation}
    \Delta J(E_{nr},J)=\delta J(E_{nr},J)+\delta J(E_{nr},-J)
\end{align}
Since the second-order contribution to $\delta J$ is odd in $J$, the above equation implies that $\Delta J$ is $\mathcal{O}(k^3)$. In particular,
\begin{align}
    \Delta J=2k^3\delta J_{3}+\mathcal{O}(k^4)
\end{align}
where, $\delta J_{3}$ in the right-hand side of the above equation is to be understood as the analytic continuation of the same (i.e., as given in \ref{delta_J_3}) to the domain $E_{nr}<0$.

\bibliography{EOM_for_EM_scattering}

\providecommand{\href}[2]{#2}\begingroup\raggedright\begin{thebibliography}{10}

\bibitem{Saketh:2021sri}
M.~V.~S. Saketh, J.~Vines, J.~Steinhoff, and A.~Buonanno, ``{Conservative and
  radiative dynamics in classical relativistic scattering and bound systems},''
  \href{http://dx.doi.org/10.1103/PhysRevResearch.4.013127}{{\em Phys. Rev.
  Res.} {\bfseries 4} no.~1, (2022) 013127},
  \href{http://arxiv.org/abs/2109.05994}{{\ttfamily arXiv:2109.05994 [gr-qc]}}.

\bibitem{LIGOScientific:2016aoc}
{\bfseries LIGO Scientific, Virgo} Collaboration, B.~P. Abbott {\em et~al.},
  ``{Observation of Gravitational Waves from a Binary Black Hole Merger},''
  \href{http://dx.doi.org/10.1103/PhysRevLett.116.061102}{{\em Phys. Rev.
  Lett.} {\bfseries 116} no.~6, (2016) 061102},
  \href{http://arxiv.org/abs/1602.03837}{{\ttfamily arXiv:1602.03837 [gr-qc]}}.

\bibitem{LIGOScientific:2016dsl}
{\bfseries LIGO Scientific, Virgo} Collaboration, B.~P. Abbott {\em et~al.},
  ``{Binary Black Hole Mergers in the first Advanced LIGO Observing Run},''
  \href{http://dx.doi.org/10.1103/PhysRevX.6.041015}{{\em Phys. Rev. X}
  {\bfseries 6} no.~4, (2016) 041015},
  \href{http://arxiv.org/abs/1606.04856}{{\ttfamily arXiv:1606.04856 [gr-qc]}}.
  [Erratum: Phys.Rev.X 8, 039903 (2018)].

\bibitem{LIGOScientific:2017vwq}
{\bfseries LIGO Scientific, Virgo} Collaboration, B.~P. Abbott {\em et~al.},
  ``{GW170817: Observation of Gravitational Waves from a Binary Neutron Star
  Inspiral},'' \href{http://dx.doi.org/10.1103/PhysRevLett.119.161101}{{\em
  Phys. Rev. Lett.} {\bfseries 119} no.~16, (2017) 161101},
  \href{http://arxiv.org/abs/1710.05832}{{\ttfamily arXiv:1710.05832 [gr-qc]}}.

\bibitem{LIGOScientific:2018mvr}
{\bfseries LIGO Scientific, Virgo} Collaboration, B.~P. Abbott {\em et~al.},
  ``{GWTC-1: A Gravitational-Wave Transient Catalog of Compact Binary Mergers
  Observed by LIGO and Virgo during the First and Second Observing Runs},''
  \href{http://dx.doi.org/10.1103/PhysRevX.9.031040}{{\em Phys. Rev. X}
  {\bfseries 9} no.~3, (2019) 031040},
  \href{http://arxiv.org/abs/1811.12907}{{\ttfamily arXiv:1811.12907
  [astro-ph.HE]}}.

\bibitem{LIGOScientific:2020ibl}
{\bfseries LIGO Scientific, Virgo} Collaboration, R.~Abbott {\em et~al.},
  ``{GWTC-2: Compact Binary Coalescences Observed by LIGO and Virgo During the
  First Half of the Third Observing Run},''
  \href{http://dx.doi.org/10.1103/PhysRevX.11.021053}{{\em Phys. Rev. X}
  {\bfseries 11} (2021) 021053},
  \href{http://arxiv.org/abs/2010.14527}{{\ttfamily arXiv:2010.14527 [gr-qc]}}.

\bibitem{Punturo:2010zz}
M.~Punturo {\em et~al.}, ``{The Einstein Telescope: A third-generation
  gravitational wave observatory},''
  \href{http://dx.doi.org/10.1088/0264-9381/27/19/194002}{{\em Class. Quant.
  Grav.} {\bfseries 27} (2010) 194002}.

\bibitem{2017arXiv170200786A}
P.~{Amaro-Seoane}, H.~{Audley}, S.~{Babak}, J.~{Baker}, E.~{Barausse},
  P.~{Bender}, E.~{Berti}, P.~{Binetruy}, M.~{Born}, D.~{Bortoluzzi},
  J.~{Camp}, C.~{Caprini}, V.~{Cardoso}, M.~{Colpi}, J.~{Conklin},
  N.~{Cornish}, C.~{Cutler}, K.~{Danzmann}, R.~{Dolesi}, L.~{Ferraioli},
  V.~{Ferroni}, E.~{Fitzsimons}, J.~{Gair}, L.~{Gesa Bote}, D.~{Giardini},
  F.~{Gibert}, C.~{Grimani}, H.~{Halloin}, G.~{Heinzel}, T.~{Hertog},
  M.~{Hewitson}, K.~{Holley-Bockelmann}, D.~{Hollington}, M.~{Hueller},
  H.~{Inchauspe}, P.~{Jetzer}, N.~{Karnesis}, C.~{Killow}, A.~{Klein},
  B.~{Klipstein}, N.~{Korsakova}, S.~L. {Larson}, J.~{Livas}, I.~{Lloro},
  N.~{Man}, D.~{Mance}, J.~{Martino}, I.~{Mateos}, K.~{McKenzie}, S.~T.
  {McWilliams}, C.~{Miller}, G.~{Mueller}, G.~{Nardini}, G.~{Nelemans},
  M.~{Nofrarias}, A.~{Petiteau}, P.~{Pivato}, E.~{Plagnol}, E.~{Porter},
  J.~{Reiche}, D.~{Robertson}, N.~{Robertson}, E.~{Rossi}, G.~{Russano},
  B.~{Schutz}, A.~{Sesana}, D.~{Shoemaker}, J.~{Slutsky}, C.~F. {Sopuerta},
  T.~{Sumner}, N.~{Tamanini}, I.~{Thorpe}, M.~{Troebs}, M.~{Vallisneri},
  A.~{Vecchio}, D.~{Vetrugno}, S.~{Vitale}, M.~{Volonteri}, G.~{Wanner},
  H.~{Ward}, P.~{Wass}, W.~{Weber}, J.~{Ziemer}, and P.~{Zweifel}, ``{Laser
  Interferometer Space Antenna},'' {\em arXiv e-prints} (Feb., 2017)
  arXiv:1702.00786, \href{http://arxiv.org/abs/1702.00786}{{\ttfamily
  arXiv:1702.00786 [astro-ph.IM]}}.

\bibitem{Reitze:2019iox}
D.~Reitze {\em et~al.}, ``{Cosmic Explorer: The U.S. Contribution to
  Gravitational-Wave Astronomy beyond LIGO},'' {\em Bull. Am. Astron. Soc.}
  {\bfseries 51} no.~7, (2019) 035,
  \href{http://arxiv.org/abs/1907.04833}{{\ttfamily arXiv:1907.04833
  [astro-ph.IM]}}.

\bibitem{Evans:2016mbw}
{\bfseries LIGO Scientific} Collaboration, B.~P. Abbott {\em et~al.},
  ``{Exploring the Sensitivity of Next Generation Gravitational Wave
  Detectors},'' \href{http://dx.doi.org/10.1088/1361-6382/aa51f4}{{\em Class.
  Quant. Grav.} {\bfseries 34} no.~4, (2017) 044001},
  \href{http://arxiv.org/abs/1607.08697}{{\ttfamily arXiv:1607.08697
  [astro-ph.IM]}}.

\bibitem{Barack:2018yly}
L.~Barack {\em et~al.}, ``{Black holes, gravitational waves and fundamental
  physics: a roadmap},'' \href{http://dx.doi.org/10.1088/1361-6382/ab0587}{{\em
  Class. Quant. Grav.} {\bfseries 36} no.~14, (2019) 143001},
  \href{http://arxiv.org/abs/1806.05195}{{\ttfamily arXiv:1806.05195 [gr-qc]}}.

\bibitem{PhysRevD.102.024060}
T.~Damour, ``Classical and quantum scattering in post-minkowskian gravity,''
  \href{http://dx.doi.org/10.1103/PhysRevD.102.024060}{{\em Phys. Rev. D}
  {\bfseries 102} (Jul, 2020) 024060}.
  \url{https://link.aps.org/doi/10.1103/PhysRevD.102.024060}.

\bibitem{PhysRevD.94.104015}
T.~Damour, ``Gravitational scattering, post-minkowskian approximation, and
  effective-one-body theory,''
  \href{http://dx.doi.org/10.1103/PhysRevD.94.104015}{{\em Phys. Rev. D}
  {\bfseries 94} (Nov, 2016) 104015}.
  \url{https://link.aps.org/doi/10.1103/PhysRevD.94.104015}.

\bibitem{PhysRevD.97.044038}
T.~Damour, ``High-energy gravitational scattering and the general relativistic
  two-body problem,'' \href{http://dx.doi.org/10.1103/PhysRevD.97.044038}{{\em
  Phys. Rev. D} {\bfseries 97} (Feb, 2018) 044038}.
  \url{https://link.aps.org/doi/10.1103/PhysRevD.97.044038}.

\bibitem{Buonanno:1998gg}
A.~Buonanno and T.~Damour, ``{Effective one-body approach to general
  relativistic two-body dynamics},''
  \href{http://dx.doi.org/10.1103/PhysRevD.59.084006}{{\em Phys. Rev. D}
  {\bfseries 59} (1999) 084006},
  \href{http://arxiv.org/abs/gr-qc/9811091}{{\ttfamily arXiv:gr-qc/9811091}}.

\bibitem{Amplitude1}
N.~E.~J. Bjerrum-Bohr, P.~H. Damgaard, G.~Festuccia, L.~Plant\'e, and
  P.~Vanhove, ``General relativity from scattering amplitudes,''
  \href{http://dx.doi.org/10.1103/PhysRevLett.121.171601}{{\em Phys. Rev.
  Lett.} {\bfseries 121} (Oct, 2018) 171601}.
  \url{https://link.aps.org/doi/10.1103/PhysRevLett.121.171601}.

\bibitem{Amplitude2}
C.~Cheung, I.~Z. Rothstein, and M.~P. Solon, ``From scattering amplitudes to
  classical potentials in the post-minkowskian expansion,''
  \href{http://dx.doi.org/10.1103/PhysRevLett.121.251101}{{\em Phys. Rev.
  Lett.} {\bfseries 121} (Dec, 2018) 251101}.
  \url{https://link.aps.org/doi/10.1103/PhysRevLett.121.251101}.

\bibitem{Neill:2013wsa}
D.~Neill and I.~Z. Rothstein, ``{Classical Space-Times from the S Matrix},''
  \href{http://dx.doi.org/10.1016/j.nuclphysb.2013.09.007}{{\em Nucl. Phys. B}
  {\bfseries 877} (2013) 177--189},
  \href{http://arxiv.org/abs/1304.7263}{{\ttfamily arXiv:1304.7263 [hep-th]}}.

\bibitem{Amplitude3}
D.~A. Kosower, B.~Maybee, and D.~O'Connell, ``{Amplitudes, Observables, and
  Classical Scattering},''
  \href{http://dx.doi.org/10.1007/JHEP02(2019)137}{{\em JHEP} {\bfseries 02}
  (2019) 137}, \href{http://arxiv.org/abs/1811.10950}{{\ttfamily
  arXiv:1811.10950 [hep-th]}}.

\bibitem{Amplitude4}
Z.~Bern, C.~Cheung, R.~Roiban, C.-H. Shen, M.~P. Solon, and M.~Zeng,
  ``Scattering amplitudes and the conservative hamiltonian for binary systems
  at third post-minkowskian order,''
  \href{http://dx.doi.org/10.1103/PhysRevLett.122.201603}{{\em Phys. Rev.
  Lett.} {\bfseries 122} (May, 2019) 201603}.
  \url{https://link.aps.org/doi/10.1103/PhysRevLett.122.201603}.

\bibitem{Amplitude5}
A.~Antonelli, A.~Buonanno, J.~Steinhoff, M.~van~de Meent, and J.~Vines,
  ``Energetics of two-body hamiltonians in post-minkowskian gravity,''
  \href{http://dx.doi.org/10.1103/PhysRevD.99.104004}{{\em Phys. Rev. D}
  {\bfseries 99} (May, 2019) 104004}.
  \url{https://link.aps.org/doi/10.1103/PhysRevD.99.104004}.

\bibitem{Amplitude6}
A.~Cristofoli, N.~E.~J. Bjerrum-Bohr, P.~H. Damgaard, and P.~Vanhove,
  ``Post-minkowskian hamiltonians in general relativity,''
  \href{http://dx.doi.org/10.1103/PhysRevD.100.084040}{{\em Phys. Rev. D}
  {\bfseries 100} (Oct, 2019) 084040}.
  \url{https://link.aps.org/doi/10.1103/PhysRevD.100.084040}.

\bibitem{Amplitude7}
G.~K\"alin and R.~A. Porto, ``{From Boundary Data to Bound States},''
  \href{http://dx.doi.org/10.1007/JHEP01(2020)072}{{\em JHEP} {\bfseries 01}
  (2020) 072}, \href{http://arxiv.org/abs/1910.03008}{{\ttfamily
  arXiv:1910.03008 [hep-th]}}.

\bibitem{Amplitude8}
N.~E.~J. Bjerrum-Bohr, A.~Cristofoli, and P.~H. Damgaard, ``{Post-Minkowskian
  Scattering Angle in Einstein Gravity},''
  \href{http://dx.doi.org/10.1007/JHEP08(2020)038}{{\em JHEP} {\bfseries 08}
  (2020) 038}, \href{http://arxiv.org/abs/1910.09366}{{\ttfamily
  arXiv:1910.09366 [hep-th]}}.

\bibitem{Amplitude9}
A.~Cristofoli, P.~H. Damgaard, P.~Di~Vecchia, and C.~Heissenberg,
  ``{Second-order Post-Minkowskian scattering in arbitrary dimensions},''
  \href{http://dx.doi.org/10.1007/JHEP07(2020)122}{{\em JHEP} {\bfseries 07}
  (2020) 122}, \href{http://arxiv.org/abs/2003.10274}{{\ttfamily
  arXiv:2003.10274 [hep-th]}}.

\bibitem{Amplitude10}
J.~Parra-Martinez, M.~S. Ruf, and M.~Zeng, ``{Extremal black hole scattering at
  $\mathcal{O}(G^3)$: graviton dominance, eikonal exponentiation, and
  differential equations},''
  \href{http://dx.doi.org/10.1007/JHEP11(2020)023}{{\em JHEP} {\bfseries 11}
  (2020) 023}, \href{http://arxiv.org/abs/2005.04236}{{\ttfamily
  arXiv:2005.04236 [hep-th]}}.

\bibitem{Amplitude11}
P.~Di~Vecchia, C.~Heissenberg, R.~Russo, and G.~Veneziano, ``{Universality of
  ultra-relativistic gravitational scattering},''
  \href{http://dx.doi.org/10.1016/j.physletb.2020.135924}{{\em Phys. Lett. B}
  {\bfseries 811} (2020) 135924},
  \href{http://arxiv.org/abs/2008.12743}{{\ttfamily arXiv:2008.12743
  [hep-th]}}.

\bibitem{Amplitude12}
T.~Damour, ``Radiative contribution to classical gravitational scattering at
  the third order in $g$,''
  \href{http://dx.doi.org/10.1103/PhysRevD.102.124008}{{\em Phys. Rev. D}
  {\bfseries 102} (Dec, 2020) 124008}.
  \url{https://link.aps.org/doi/10.1103/PhysRevD.102.124008}.

\bibitem{Amplitude13}
P.~Di~Vecchia, C.~Heissenberg, R.~Russo, and G.~Veneziano, ``{Radiation
  Reaction from Soft Theorems},''
  \href{http://dx.doi.org/10.1016/j.physletb.2021.136379}{{\em Phys. Lett. B}
  {\bfseries 818} (2021) 136379},
  \href{http://arxiv.org/abs/2101.05772}{{\ttfamily arXiv:2101.05772
  [hep-th]}}.

\bibitem{Amplitude14}
Z.~Bern, J.~Parra-Martinez, R.~Roiban, M.~S. Ruf, C.-H. Shen, M.~P. Solon, and
  M.~Zeng, ``Scattering amplitudes and conservative binary dynamics at
  $\mathcal{O}({G}^{4})$,''
  \href{http://dx.doi.org/10.1103/PhysRevLett.126.171601}{{\em Phys. Rev.
  Lett.} {\bfseries 126} (Apr, 2021) 171601}.
  \url{https://link.aps.org/doi/10.1103/PhysRevLett.126.171601}.

\bibitem{Amplitude15}
P.~Di~Vecchia, C.~Heissenberg, R.~Russo, and G.~Veneziano, ``{The eikonal
  approach to gravitational scattering and radiation at $ \mathcal{O}
  $(G$^{3}$)},'' \href{http://dx.doi.org/10.1007/JHEP07(2021)169}{{\em JHEP}
  {\bfseries 07} (2021) 169}, \href{http://arxiv.org/abs/2104.03256}{{\ttfamily
  arXiv:2104.03256 [hep-th]}}.

\bibitem{Amplitude16}
N.~E.~J. Bjerrum-Bohr, P.~H. Damgaard, L.~Plant\'e, and P.~Vanhove, ``Classical
  gravity from loop amplitudes,''
  \href{http://dx.doi.org/10.1103/PhysRevD.104.026009}{{\em Phys. Rev. D}
  {\bfseries 104} (Jul, 2021) 026009}.
  \url{https://link.aps.org/doi/10.1103/PhysRevD.104.026009}.

\bibitem{Amplitude17}
N.~E.~J. Bjerrum-Bohr, P.~H. Damgaard, L.~Plant\'e, and P.~Vanhove, ``{The
  amplitude for classical gravitational scattering at third Post-Minkowskian
  order},'' \href{http://dx.doi.org/10.1007/JHEP08(2021)172}{{\em JHEP}
  {\bfseries 08} (2021) 172}, \href{http://arxiv.org/abs/2105.05218}{{\ttfamily
  arXiv:2105.05218 [hep-th]}}.

\bibitem{Amplitude18}
D.~Bini, T.~Damour, and A.~Geralico, ``Radiative contributions to gravitational
  scattering,'' \href{http://dx.doi.org/10.1103/PhysRevD.104.084031}{{\em Phys.
  Rev. D} {\bfseries 104} (Oct, 2021) 084031}.
  \url{https://link.aps.org/doi/10.1103/PhysRevD.104.084031}.

\bibitem{Amplitude19}
Y.~F. Bautista, A.~Guevara, C.~Kavanagh, and J.~Vines, ``{From Scattering in
  Black Hole Backgrounds to Higher-Spin Amplitudes: Part I},''
  \href{http://arxiv.org/abs/2107.10179}{{\ttfamily arXiv:2107.10179
  [hep-th]}}.

\bibitem{Amplitude20}
A.~Cristofoli, R.~Gonzo, D.~A. Kosower, and D.~O'Connell, ``{Waveforms from
  amplitudes},'' \href{http://dx.doi.org/10.1103/PhysRevD.106.056007}{{\em
  Phys. Rev. D} {\bfseries 106} no.~5, (2022) 056007},
  \href{http://arxiv.org/abs/2107.10193}{{\ttfamily arXiv:2107.10193
  [hep-th]}}.

\bibitem{Amplitude21}
E.~Herrmann, J.~Parra-Martinez, M.~S. Ruf, and M.~Zeng, ``Gravitational
  bremsstrahlung from reverse unitarity,''
  \href{http://dx.doi.org/10.1103/PhysRevLett.126.201602}{{\em Phys. Rev.
  Lett.} {\bfseries 126} (May, 2021) 201602}.
  \url{https://link.aps.org/doi/10.1103/PhysRevLett.126.201602}.

\bibitem{Amplitude22}
E.~Herrmann, J.~Parra-Martinez, M.~S. Ruf, and M.~Zeng, ``{Radiative classical
  gravitational observables at $ \mathcal{O} $(G$^{3}$) from scattering
  amplitudes},'' \href{http://dx.doi.org/10.1007/JHEP10(2021)148}{{\em JHEP}
  {\bfseries 10} (2021) 148}, \href{http://arxiv.org/abs/2104.03957}{{\ttfamily
  arXiv:2104.03957 [hep-th]}}.

\bibitem{Amplitude23}
S.~Mougiakakos, M.~M. Riva, and F.~Vernizzi, ``Gravitational bremsstrahlung in
  the post-minkowskian effective field theory,''
  \href{http://dx.doi.org/10.1103/PhysRevD.104.024041}{{\em Phys. Rev. D}
  {\bfseries 104} (Jul, 2021) 024041}.
  \url{https://link.aps.org/doi/10.1103/PhysRevD.104.024041}.

\bibitem{Amplitude24}
G.~U. Jakobsen, G.~Mogull, J.~Plefka, and J.~Steinhoff, ``Classical
  gravitational bremsstrahlung from a worldline quantum field theory,''
  \href{http://dx.doi.org/10.1103/PhysRevLett.126.201103}{{\em Phys. Rev.
  Lett.} {\bfseries 126} (May, 2021) 201103}.
  \url{https://link.aps.org/doi/10.1103/PhysRevLett.126.201103}.

\bibitem{Amplitude25}
P.~H. Damgaard, L.~Plante, and P.~Vanhove, ``{On an exponential representation
  of the gravitational S-matrix},''
  \href{http://dx.doi.org/10.1007/JHEP11(2021)213}{{\em JHEP} {\bfseries 11}
  (2021) 213}, \href{http://arxiv.org/abs/2107.12891}{{\ttfamily
  arXiv:2107.12891 [hep-th]}}.

\bibitem{Amplitude26}
A.~Brandhuber, G.~Chen, G.~Travaglini, and C.~Wen, ``{Classical gravitational
  scattering from a gauge-invariant double copy},''
  \href{http://dx.doi.org/10.1007/JHEP10(2021)118}{{\em JHEP} {\bfseries 10}
  (2021) 118}, \href{http://arxiv.org/abs/2108.04216}{{\ttfamily
  arXiv:2108.04216 [hep-th]}}.

\bibitem{Kalin:2019inp}
G.~K\"alin and R.~A. Porto, ``{From boundary data to bound states. Part II.
  Scattering angle to dynamical invariants (with twist)},''
  \href{http://dx.doi.org/10.1007/JHEP02(2020)120}{{\em JHEP} {\bfseries 02}
  (2020) 120}, \href{http://arxiv.org/abs/1911.09130}{{\ttfamily
  arXiv:1911.09130 [hep-th]}}.

\bibitem{Damgaard:2021rnk}
P.~H. Damgaard and P.~Vanhove, ``{Remodeling the effective one-body formalism
  in post-Minkowskian gravity},''
  \href{http://dx.doi.org/10.1103/PhysRevD.104.104029}{{\em Phys. Rev. D}
  {\bfseries 104} no.~10, (2021) 104029},
  \href{http://arxiv.org/abs/2108.11248}{{\ttfamily arXiv:2108.11248
  [hep-th]}}.

\bibitem{Manohar:2022dea}
A.~V. Manohar, A.~K. Ridgway, and C.-H. Shen, ``{Radiated Angular Momentum and
  Dissipative Effects in Classical Scattering},''
  \href{http://dx.doi.org/10.1103/PhysRevLett.129.121601}{{\em Phys. Rev.
  Lett.} {\bfseries 129} no.~12, (2022) 121601},
  \href{http://arxiv.org/abs/2203.04283}{{\ttfamily arXiv:2203.04283
  [hep-th]}}.

\bibitem{Dlapa:2022lmu}
C.~Dlapa, G.~K\"alin, Z.~Liu, J.~Neef, and R.~A. Porto, ``{Radiation Reaction
  and Gravitational Waves at Fourth Post-Minkowskian Order},''
  \href{http://arxiv.org/abs/2210.05541}{{\ttfamily arXiv:2210.05541
  [hep-th]}}.

\bibitem{Damour:2020tta}
T.~Damour, ``{Radiative contribution to classical gravitational scattering at
  the third order in $G$},''
  \href{http://dx.doi.org/10.1103/PhysRevD.102.124008}{{\em Phys. Rev. D}
  {\bfseries 102} no.~12, (2020) 124008},
  \href{http://arxiv.org/abs/2010.01641}{{\ttfamily arXiv:2010.01641 [gr-qc]}}.

\bibitem{Cho:2021arx}
G.~Cho, G.~K\"alin, and R.~A. Porto, ``{From boundary data to bound states.
  Part III. Radiative effects},''
  \href{http://dx.doi.org/10.1007/JHEP04(2022)154}{{\em JHEP} {\bfseries 04}
  (2022) 154}, \href{http://arxiv.org/abs/2112.03976}{{\ttfamily
  arXiv:2112.03976 [hep-th]}}. [Erratum: JHEP 07, 002 (2022)].

\bibitem{Bini:2021qvf}
D.~Bini and A.~Geralico, ``{Higher-order tail contributions to the energy and
  angular momentum fluxes in a two-body scattering process},''
  \href{http://dx.doi.org/10.1103/PhysRevD.104.104020}{{\em Phys. Rev. D}
  {\bfseries 104} no.~10, (2021) 104020},
  \href{http://arxiv.org/abs/2108.05445}{{\ttfamily arXiv:2108.05445 [gr-qc]}}.

\bibitem{Kawai:1985xq}
H.~Kawai, D.~C. Lewellen, and S.~H.~H. Tye, ``{A Relation Between Tree
  Amplitudes of Closed and Open Strings},''
  \href{http://dx.doi.org/10.1016/0550-3213(86)90362-7}{{\em Nucl. Phys. B}
  {\bfseries 269} (1986) 1--23}.

\bibitem{Bern:2008qj}
Z.~Bern, J.~J.~M. Carrasco, and H.~Johansson, ``{New Relations for Gauge-Theory
  Amplitudes},'' \href{http://dx.doi.org/10.1103/PhysRevD.78.085011}{{\em Phys.
  Rev. D} {\bfseries 78} (2008) 085011},
  \href{http://arxiv.org/abs/0805.3993}{{\ttfamily arXiv:0805.3993 [hep-ph]}}.

\bibitem{Bern:2010ue}
Z.~Bern, J.~J.~M. Carrasco, and H.~Johansson, ``{Perturbative Quantum Gravity
  as a Double Copy of Gauge Theory},''
  \href{http://dx.doi.org/10.1103/PhysRevLett.105.061602}{{\em Phys. Rev.
  Lett.} {\bfseries 105} (2010) 061602},
  \href{http://arxiv.org/abs/1004.0476}{{\ttfamily arXiv:1004.0476 [hep-th]}}.

\bibitem{Bern:2019prr}
Z.~Bern, J.~J. Carrasco, M.~Chiodaroli, H.~Johansson, and R.~Roiban, ``{The
  Duality Between Color and Kinematics and its Applications},''
  \href{http://arxiv.org/abs/1909.01358}{{\ttfamily arXiv:1909.01358
  [hep-th]}}.

\bibitem{Damour:2022ybd}
T.~Damour and P.~Rettegno, ``{Strong-field scattering of two black holes:
  Numerical Relativity meets Post-Minkowskian gravity},''
  \href{http://arxiv.org/abs/2211.01399}{{\ttfamily arXiv:2211.01399 [gr-qc]}}.

\bibitem{Westpfahl:1985tsl}
K.~Westpfahl, ``{High-Speed Scattering of Charged and Uncharged Particles in
  General Relativity},'' \href{http://dx.doi.org/10.1002/prop.2190330802}{{\em
  Fortsch. Phys.} {\bfseries 33} no.~8, (1985) 417--493}.

\bibitem{Buonanno:2000ef}
A.~Buonanno and T.~Damour, ``{Transition from inspiral to plunge in binary
  black hole coalescences},''
  \href{http://dx.doi.org/10.1103/PhysRevD.62.064015}{{\em Phys. Rev. D}
  {\bfseries 62} (2000) 064015},
  \href{http://arxiv.org/abs/gr-qc/0001013}{{\ttfamily arXiv:gr-qc/0001013}}.

\bibitem{Damour:2000we}
T.~Damour, P.~Jaranowski, and G.~Schaefer, ``{On the determination of the last
  stable orbit for circular general relativistic binaries at the third
  postNewtonian approximation},''
  \href{http://dx.doi.org/10.1103/PhysRevD.62.084011}{{\em Phys. Rev. D}
  {\bfseries 62} (2000) 084011},
  \href{http://arxiv.org/abs/gr-qc/0005034}{{\ttfamily arXiv:gr-qc/0005034}}.

\bibitem{Damour:2001tu}
T.~Damour, ``{Coalescence of two spinning black holes: an effective one-body
  approach},'' \href{http://dx.doi.org/10.1103/PhysRevD.64.124013}{{\em Phys.
  Rev. D} {\bfseries 64} (2001) 124013},
  \href{http://arxiv.org/abs/gr-qc/0103018}{{\ttfamily arXiv:gr-qc/0103018}}.

\bibitem{Buonanno:2005xu}
A.~Buonanno, Y.~Chen, and T.~Damour, ``{Transition from inspiral to plunge in
  precessing binaries of spinning black holes},''
  \href{http://dx.doi.org/10.1103/PhysRevD.74.104005}{{\em Phys. Rev. D}
  {\bfseries 74} (2006) 104005},
  \href{http://arxiv.org/abs/gr-qc/0508067}{{\ttfamily arXiv:gr-qc/0508067}}.

\bibitem{PhysRevD.1.2349}
E.~Brezin, C.~Itzykson, and J.~Zinn-Justin, ``Relativistic balmer formula
  including recoil effects,''
  \href{http://dx.doi.org/10.1103/PhysRevD.1.2349}{{\em Phys. Rev. D}
  {\bfseries 1} (Apr, 1970) 2349--2355}.
  \url{https://link.aps.org/doi/10.1103/PhysRevD.1.2349}.

\bibitem{PhysRevD.73.104029}
W.~D. Goldberger and I.~Z. Rothstein, ``Effective field theory of gravity for
  extended objects,'' \href{http://dx.doi.org/10.1103/PhysRevD.73.104029}{{\em
  Phys. Rev. D} {\bfseries 73} (May, 2006) 104029}.
  \url{https://link.aps.org/doi/10.1103/PhysRevD.73.104029}.

\bibitem{Goldberger:2006bd}
W.~D. Goldberger and I.~Z. Rothstein, ``{Towers of Gravitational Theories},''
  \href{http://dx.doi.org/10.1142/S0218271806009698}{{\em Gen. Rel. Grav.}
  {\bfseries 38} (2006) 1537--1546},
  \href{http://arxiv.org/abs/hep-th/0605238}{{\ttfamily arXiv:hep-th/0605238}}.

\bibitem{PhysRevD.73.104031}
R.~A. Porto, ``Post-newtonian corrections to the motion of spinning bodies in
  nonrelativistic general relativity,''
  \href{http://dx.doi.org/10.1103/PhysRevD.73.104031}{{\em Phys. Rev. D}
  {\bfseries 73} (May, 2006) 104031}.
  \url{https://link.aps.org/doi/10.1103/PhysRevD.73.104031}.

\bibitem{porto2007effective}
R.~A. Porto, {\em An effective field theory of gravity for spinning extended
  objects}.
\newblock PhD thesis, Carnegie Mellon University, 2007.

\bibitem{Porto:2016pyg}
R.~A. Porto, ``{The effective field theorist\textquoteright{}s approach to
  gravitational dynamics},''
  \href{http://dx.doi.org/10.1016/j.physrep.2016.04.003}{{\em Phys. Rept.}
  {\bfseries 633} (2016) 1--104},
  \href{http://arxiv.org/abs/1601.04914}{{\ttfamily arXiv:1601.04914
  [hep-th]}}.

\bibitem{Kalin:2020mvi}
G.~K\"alin and R.~A. Porto, ``{Post-Minkowskian Effective Field Theory for
  Conservative Binary Dynamics},''
  \href{http://dx.doi.org/10.1007/JHEP11(2020)106}{{\em JHEP} {\bfseries 11}
  (2020) 106}, \href{http://arxiv.org/abs/2006.01184}{{\ttfamily
  arXiv:2006.01184 [hep-th]}}.

\bibitem{Kalin:2020fhe}
G.~K\"alin, Z.~Liu, and R.~A. Porto, ``{Conservative Dynamics of Binary Systems
  to Third Post-Minkowskian Order from the Effective Field Theory Approach},''
  \href{http://dx.doi.org/10.1103/PhysRevLett.125.261103}{{\em Phys. Rev.
  Lett.} {\bfseries 125} no.~26, (2020) 261103},
  \href{http://arxiv.org/abs/2007.04977}{{\ttfamily arXiv:2007.04977
  [hep-th]}}.

\bibitem{Kalin:2020lmz}
G.~K\"alin, Z.~Liu, and R.~A. Porto, ``{Conservative Tidal Effects in Compact
  Binary Systems to Next-to-Leading Post-Minkowskian Order},''
  \href{http://dx.doi.org/10.1103/PhysRevD.102.124025}{{\em Phys. Rev. D}
  {\bfseries 102} (2020) 124025},
  \href{http://arxiv.org/abs/2008.06047}{{\ttfamily arXiv:2008.06047
  [hep-th]}}.

\bibitem{Liu:2021zxr}
Z.~Liu, R.~A. Porto, and Z.~Yang, ``{Spin Effects in the Effective Field Theory
  Approach to Post-Minkowskian Conservative Dynamics},''
  \href{http://dx.doi.org/10.1007/JHEP06(2021)012}{{\em JHEP} {\bfseries 06}
  (2021) 012}, \href{http://arxiv.org/abs/2102.10059}{{\ttfamily
  arXiv:2102.10059 [hep-th]}}.

\bibitem{Dlapa:2021npj}
C.~Dlapa, G.~K\"alin, Z.~Liu, and R.~A. Porto, ``{Dynamics of binary systems to
  fourth Post-Minkowskian order from the effective field theory approach},''
  \href{http://dx.doi.org/10.1016/j.physletb.2022.137203}{{\em Phys. Lett. B}
  {\bfseries 831} (2022) 137203},
  \href{http://arxiv.org/abs/2106.08276}{{\ttfamily arXiv:2106.08276
  [hep-th]}}.

\bibitem{Dlapa:2021vgp}
C.~Dlapa, G.~K\"alin, Z.~Liu, and R.~A. Porto, ``{Conservative Dynamics of
  Binary Systems at Fourth Post-Minkowskian Order in the Large-Eccentricity
  Expansion},'' \href{http://dx.doi.org/10.1103/PhysRevLett.128.161104}{{\em
  Phys. Rev. Lett.} {\bfseries 128} no.~16, (2022) 161104},
  \href{http://arxiv.org/abs/2112.11296}{{\ttfamily arXiv:2112.11296
  [hep-th]}}.

\bibitem{Kalin:2022hph}
G.~K\"alin, J.~Neef, and R.~A. Porto, ``{Radiation-Reaction in the Effective
  Field Theory Approach to Post-Minkowskian Dynamics},''
  \href{http://arxiv.org/abs/2207.00580}{{\ttfamily arXiv:2207.00580
  [hep-th]}}.

\bibitem{Jakobsen:2022psy}
G.~U. Jakobsen, G.~Mogull, J.~Plefka, and B.~Sauer, ``{All things retarded:
  radiation-reaction in worldline quantum field theory},''
  \href{http://dx.doi.org/10.1007/JHEP10(2022)128}{{\em JHEP} {\bfseries 10}
  (2022) 128}, \href{http://arxiv.org/abs/2207.00569}{{\ttfamily
  arXiv:2207.00569 [hep-th]}}.

\bibitem{Jakobsen:2021smu}
G.~U. Jakobsen, G.~Mogull, J.~Plefka, and J.~Steinhoff, ``{Classical
  Gravitational Bremsstrahlung from a Worldline Quantum Field Theory},''
  \href{http://dx.doi.org/10.1103/PhysRevLett.126.201103}{{\em Phys. Rev.
  Lett.} {\bfseries 126} no.~20, (2021) 201103},
  \href{http://arxiv.org/abs/2101.12688}{{\ttfamily arXiv:2101.12688 [gr-qc]}}.

\bibitem{Jakobsen:2022fcj}
G.~U. Jakobsen and G.~Mogull, ``{Conservative and Radiative Dynamics of
  Spinning Bodies at Third Post-Minkowskian Order Using Worldline Quantum Field
  Theory},'' \href{http://dx.doi.org/10.1103/PhysRevLett.128.141102}{{\em Phys.
  Rev. Lett.} {\bfseries 128} no.~14, (2022) 141102},
  \href{http://arxiv.org/abs/2201.07778}{{\ttfamily arXiv:2201.07778
  [hep-th]}}.

\bibitem{Mougiakakos:2021ckm}
S.~Mougiakakos, M.~M. Riva, and F.~Vernizzi, ``{Gravitational Bremsstrahlung in
  the post-Minkowskian effective field theory},''
  \href{http://dx.doi.org/10.1103/PhysRevD.104.024041}{{\em Phys. Rev. D}
  {\bfseries 104} no.~2, (2021) 024041},
  \href{http://arxiv.org/abs/2102.08339}{{\ttfamily arXiv:2102.08339 [gr-qc]}}.

\bibitem{Riva:2021vnj}
M.~M. Riva and F.~Vernizzi, ``{Radiated momentum in the post-Minkowskian
  worldline approach via reverse unitarity},''
  \href{http://dx.doi.org/10.1007/JHEP11(2021)228}{{\em JHEP} {\bfseries 11}
  (2021) 228}, \href{http://arxiv.org/abs/2110.10140}{{\ttfamily
  arXiv:2110.10140 [hep-th]}}.

\bibitem{Riva:2022fru}
M.~M. Riva, F.~Vernizzi, and L.~K. Wong, ``{Gravitational bremsstrahlung from
  spinning binaries in the post-Minkowskian expansion},''
  \href{http://dx.doi.org/10.1103/PhysRevD.106.044013}{{\em Phys. Rev. D}
  {\bfseries 106} no.~4, (2022) 044013},
  \href{http://arxiv.org/abs/2205.15295}{{\ttfamily arXiv:2205.15295
  [hep-th]}}.

\bibitem{Bini:2012ji}
D.~Bini and T.~Damour, ``{Gravitational radiation reaction along general orbits
  in the effective one-body formalism},''
  \href{http://dx.doi.org/10.1103/PhysRevD.86.124012}{{\em Phys. Rev. D}
  {\bfseries 86} (2012) 124012},
  \href{http://arxiv.org/abs/1210.2834}{{\ttfamily arXiv:1210.2834 [gr-qc]}}.

\bibitem{PhysRevLett.110.174301}
C.~R. Galley, ``Classical mechanics of nonconservative systems,''
  \href{http://dx.doi.org/10.1103/PhysRevLett.110.174301}{{\em Phys. Rev.
  Lett.} {\bfseries 110} (Apr, 2013) 174301}.
  \url{https://link.aps.org/doi/10.1103/PhysRevLett.110.174301}.

\bibitem{PhysRevD.55.3457}
Y.~Mino, M.~Sasaki, and T.~Tanaka, ``Gravitational radiation reaction to a
  particle motion,'' \href{http://dx.doi.org/10.1103/PhysRevD.55.3457}{{\em
  Phys. Rev. D} {\bfseries 55} (Mar, 1997) 3457--3476}.
  \url{https://link.aps.org/doi/10.1103/PhysRevD.55.3457}.

\bibitem{PhysRevD.56.3381}
T.~C. Quinn and R.~M. Wald, ``Axiomatic approach to electromagnetic and
  gravitational radiation reaction of particles in curved spacetime,''
  \href{http://dx.doi.org/10.1103/PhysRevD.56.3381}{{\em Phys. Rev. D}
  {\bfseries 56} (Sep, 1997) 3381--3394}.
  \url{https://link.aps.org/doi/10.1103/PhysRevD.56.3381}.

\bibitem{Landau:1975pou}
L.~D. Landau and E.~M. Lifschits, {\em {The Classical Theory of Fields}},
  vol.~Volume 2 of {\em Course of Theoretical Physics}.
\newblock Pergamon Press, Oxford, 1975.

\bibitem{Bini:2020hmy}
D.~Bini, T.~Damour, and A.~Geralico, ``{Sixth post-Newtonian nonlocal-in-time
  dynamics of binary systems},''
  \href{http://dx.doi.org/10.1103/PhysRevD.102.084047}{{\em Phys. Rev. D}
  {\bfseries 102} no.~8, (2020) 084047},
  \href{http://arxiv.org/abs/2007.11239}{{\ttfamily arXiv:2007.11239 [gr-qc]}}.

\end{thebibliography}\endgroup

\bibliographystyle{utphys1}
\end{document}